\documentclass[a4paper,12pt]{article}
\pdfoutput=1
\usepackage{graphicx,rotating,amsmath,amsfonts,amssymb}
\usepackage{hyperref}\usepackage{slashed}
\hypersetup{colorlinks,bookmarksopen,bookmarksnumbered,
linkcolor=blus,pdfstartview=FitH,urlcolor=rossos,citecolor=verde}

\allowdisplaybreaks
\usepackage{multicol}
\usepackage{color}
\definecolor{rosso}{cmyk}{0,1,1,0.4}
\definecolor{rossos}{cmyk}{0,1,1,0.55}
\definecolor{rossoc}{cmyk}{0,1,1,0.2}
\definecolor{blu}{cmyk}{1,1,0,0.3}
\definecolor{blus}{cmyk}{1,1,0,0.6}
\definecolor{bluc}{cmyk}{1,1,0,0.1}
\definecolor{verde}{cmyk}{0.92,0,0.59,0.25}
\definecolor{verdec}{cmyk}{0.92,0,0.59,0.15}
\definecolor{verdes}{cmyk}{0.92,0,0.59,0.4}

\newcommand{\PRexp}{2.14 \pm 0.05}
\newcommand{\bp}{\bar M_{\rm Pl}}
\newcommand{\mub}{\bar{\mu}}

\newcommand{\eq}[1]{~{\rm (\ref{eq:#1})}}

\newcommand{\rfn}[1]{(\ref{#1})}

\newcommand{\GeV}{\,{\rm GeV}}
\newcommand{\TeV}{\,{\rm TeV}}

\def\circa#1{\,\raise.3ex\hbox{$#1$\kern-.75em\lower1ex\hbox{$\sim$}}\,}

\newcommand{\beq}{\begin{equation}}
\newcommand{\eeq}{\end{equation}}

\newcommand{\bea}{\begin{eqnarray}}
\newcommand{\eea}{\end{eqnarray}}
\newcommand{\be}{\begin{equation}}
\newcommand{\ee}{\end{equation}}
\font\tenrsfs=rsfs10 at 12pt
\font\sevenrsfs=rsfs7
\font\fiversfs=rsfs5
\newfam\rsfsfam
\textfont\rsfsfam=\tenrsfs
\scriptfont\rsfsfam=\sevenrsfs
\scriptscriptfont\rsfsfam=\fiversfs
\def\mathscr#1{{\fam\rsfsfam\relax#1}}
\newcommand{\gsim}{\lower.7ex\hbox{$\;\stackrel{\textstyle>}{\sim}\;$}}
\newcommand{\lsim}{\lower.7ex\hbox{$\;\stackrel{\textstyle<}{\sim}\;$}}

\def\Lag{\mathscr{L}}

\def\circa#1{\,\raise.3ex\hbox{$#1$\kern-.75em\lower1ex\hbox{$\sim$}}\,}
\makeatletter

\def\art{\@ifnextchar[{\eart}{\oart}}
\def\eart[#1]#2#3#4#5#6{{#2}, {#3  #4} {\rm (#6) #5} ({\em #1})}
\def\hepart[#1]#2{{\rm #2, #1}}
\newcommand{\oart}[5]{{\rm #1}, {#2 #3} {\rm (#5) #4}}

\def\hhref#1{\href{http://arxiv.org/abs/#1}{arXiv:#1}} 
\def\arXiv#1{\href{http://arxiv.org/abs/#1}{arXiv:#1}} 
\newcounter{alphaequation}[equation]
\def\thealphaequation{\theequation\hbox to
0.6em{\hfil\alph{alphaequation}\hfil}}
\def\eqnsystem#1{
\def\@eqnnum{{\rm (\thealphaequation)}}
\def\@@eqncr{\let\@tempa\relax \ifcase\@eqcnt \def\@tempa{& & &} \or
  \def\@tempa{& &}\or \def\@tempa{&}\fi\@tempa
  \if@eqnsw\@eqnnum\refstepcounter{alphaequation}\fi
\global\@eqnswtrue\global\@eqcnt=0\cr}
\refstepcounter{equation} \let\@currentlabel\theequation \def\@tempb{#1}
\ifx\@tempb\empty\else\label{#1}\fi
\refstepcounter{alphaequation}
\let\@currentlabel\thealphaequation
\global\@eqnswtrue\global\@eqcnt=0 \tabskip\@centering\let\\=\@eqncr
$$\halign to \displaywidth\bgroup \@eqnsel\hskip\@centering
$\displaystyle\tabskip\z@{##}$&\global\@eqcnt\@ne
\hskip2\arraycolsep\hfil${##}$\hfil& \global\@eqcnt\tw@\hskip2\arraycolsep
$\displaystyle\tabskip\z@{##}$\hfil
\tabskip\@centering&\llap{##}\tabskip\z@\cr}
\def\endeqnsystem{\@@eqncr\egroup$$\global\@ignoretrue} \makeatother

\newcommand{\eV}{\,{\rm eV}}

\begin{document}
\begin{center}
 IFUP-TH/2015  \hspace{2cm} IFT-UAM/CSIC-15-015\\
\vspace*{0.8cm}
{\LARGE\Huge\bf\color{rossos} Dynamically Induced\\[1mm] Planck Scale and Inflation}\\

\bigskip
\bigskip
\bigskip

{\bf Kristjan Kannike$^a$, Gert H\"utsi$^b$, Liberato Pizza$^c$,
Antonio Racioppi$^a$,  \\[1mm]
Martti Raidal$^{a,d}$, Alberto Salvio$^e$
{\rm and}
Alessandro Strumia$^{a,c}$}
\\[7mm]

{
{\it $^a$ NICPB, R\"avala 10, 10143  Tallinn, Estonia}

{\it $^{b}$ Tartu Observatory, Observatooriumi 1, 61602 T\~oravere, Estonia}

{\it $^c$ Dipartimento di Fisica dell'Universit{\`a} di Pisa and INFN, Italy}

{\it $^d$ Institute of Physics, University of Tartu, Ravila 14c, 50411 Tartu, Estonia}

{\it $^e$ Departamento de F\'isica Te\'orica, Universidad Aut\'onoma de Madrid\\ and Instituto de F\'isica Te\'orica IFT-UAM/CSIC,  Madrid, Spain}

}

\vspace{1cm}\large
{\large\bf\color{blus} Abstract}
\begin{quote}
Theories where the Planck scale is dynamically generated from dimensionless interactions provide predictive inflationary potentials and super-Planckian field variations.
We first study the minimal single field realisation in the low-energy effective field theory limit,
finding the predictions  $n_s \approx 0.96$  for the spectral index and $r \approx 0.13$ for the tensor-to-scalar ratio,
which can be reduced down to $\approx 0.04$ in presence of large couplings.
Next we consider agravity as a dimensionless quantum gravity theory finding a multifield inflation that converges  towards an attractor trajectory and predicts
$n_s\approx 0.96$ and $0.003<r<0.13$, interpolating between the quadratic and Starobinsky inflation.
These theories relate the  smallness of the weak scale  to the smallness of inflationary perturbations:
both arise naturally because of small couplings, implying a reheating temperature of $10^{7-9}$~GeV.
A measurement of $r$ by {\sc Keck/Bicep3} would give us  information on  quantum gravity in the dimensionless scenario.

\end{quote}
\thispagestyle{empty}
\end{center}

\newpage
\tableofcontents

\section{Introduction}

The discovery of the Higgs boson~\cite{Chatrchyan:2012ufa,Aad:2012tfa}
and the lack (so far) of new physics challenged the standard view on naturalness of the electroweak scale~\cite{naturalnessconf}.
The latter led to the expectation that the Higgs boson should be accompanied by new physics at the weak scale that is able to
provide  a cut-off to quadratically divergent quantum corrections to the  squared Higgs mass due to the standard model (SM) couplings.

Given that power divergent quantum corrections do not lead to any physical effect,
some theorists are considering the possibility that the Higgs mass fine-tuning problem
could be just an unphysical artifact of the standard renormalisation procedure,
that introduces an artificial cut-off and unphysical bare parameters.
On the contrary, heavy new particles coupled to the Higgs boson would lead to large physical corrections to the Higgs mass:
the associated fine-tuning could be probed by experiments~\cite{FN}.
Thereby, one is led to redefine natural models as those where new physics heavier than the weak scale is weakly coupled to the Higgs.

Starting from these phenomenological considerations, various authors tried to develop a theoretical framework able of
explaining the co-existence and the origin of the largely separated mass scales observed in nature.
Most attempts involve, in some way or another,  classical scale invariance and dynamical generation of mass scales~\cite{FNmodels,agravity}.
Implications for inflation have been explored in~\cite{agravity,Kannike:2014mia}.

The largest scale observed in Nature so far  is the Planck scale. Inflationary~\cite{Guth:1980zm} generation of primordial perturbations~\cite{Mukhanov:1981xt}
 can provide an observational window on Planck-scale physics.
The recent {\sc Bicep2/Keck/Planck} common analysis~\cite{BICEP2/Keck:2015tva} of the $B$-mode polarisation data
tries to control the astrophysical backgrounds~\cite{Adam:2014bub}  and
hints at a  value for the tensor-to-scalar ratio $r=0.06\pm0.04$,
in-between the previous
BICEP2~\cite{Ade:2014xna} and Planck~\cite{Ade:2013uln,Planck2015}  results.
If future experiments will find a statistically significant evidence for $r>0$,
this might be the first hint of the quantum nature of gravity, and a precise determination of $r$
may help us to discriminate between different ultraviolet (UV) completions of gravity.\footnote{However, an observable value of $r$  can also be obtained  for sub-Planckian field variations in certain cases \cite{Hotchkiss:2011gz}.} This result, therefore, invites for thorough common studies of gravity and inflation.

In this paper we explore the implications of the assumption that the Planck scale is generated dynamically,
assuming that the same sector also provides inflation.
The dynamics leading to
dimensional transmutation can be due to strongly-coupled or to weakly-coupled physics.
We here focus on weakly coupled dynamics, such that we can perform perturbative computations.
The literature contains studies of models with a dynamical Planck scale (see e.g.~\cite{cosmon,agravity})
and of models with a dynamical inflaton potential
(already the very first papers on inflation considered the possibility that the inflaton potential is generated dynamically by loop effects~\cite{Ellis:1982dg} via
the Coleman-Weinberg mechanism~\cite{Coleman:1973jx}).
Furthermore, dynamical generation of masses is compatible with the small observed cosmological constant provided
that the scalar potential satisfies the `multiple point criticality principle'~\cite{Froggatt:1995rt}, that was  introduced for the SM
Higgs boson and is extensively used in Higgs inflation~\cite{Bezrukov:2007ep,Bezrukov:2010jz}. Originally, the non-minimal coupling for the usual inflaton was considered in \cite{orig:xi}.

In this paper we combine these concepts into a consistent framework and study implications for inflation,
gravity and the electroweak scale.  We follow two different approaches.

\medskip

To obtain model independent results we first take an effective field theory approach and study the minimal single field inflation from dynamical
generation of the Planck scale without knowing the theory of gravity.
We assume that  a complete theory of quantum gravity is not needed either because inflation is described by Einstein gravity at sub-Planckian energy
or because some completion of Einstein gravity is  weakly coupled enough.
The inflaton is assumed to be the Higgs of gravity: the pseudo-Goldstone boson
of scale invariance that acquires a vacuum expectation value (VEV) generating the Planck mass.
The assumption of classical scale invariance allows us to deal with trans-Planckian inflaton field values~\cite{Lyth:1996im}.

 We find that in the limit when gravity effects can be ignored the inflationary observables converge towards the predictions of a  quadratic inflationary
potential~\cite{Linde:1983gd}, up to deviations due to higher order corrections, $n_s \approx 0.96$ and $r \approx 0.13^{+0.01}_{-0.03}$.
We formulate conditions when this approximation is valid (details are collected in appendix~\ref{app})
and discuss the equivalence of Einstein and Jordan frames in this limit.
For large values of the non-minimal coupling to gravity, we obtain much wider range for tensor-to-scalar ratio, $r>0.04,$ while the
prediction for scalar spectral index remains the same.
This is new and much more constraining result compared to the corresponding result in the previous scale-invariant inflation study~\cite{Kannike:2014mia}
due to extra constraints arising from the dynamically generated Planck scale and the dynamically realised multiple point criticality principle.
We present the minimal model for this scenario and study its properties.

\medskip

After the effective field theory study, we focus on a specific possibility for quantum gravity:
agravity~\cite{agravity}, which is the dimensionless renormalizable extension of Einstein gravity.
New gravitational degrees of freedom,  predicted by the theory, can be light enough to take part in inflationary dynamics.
We thereby have multifield inflation, and we find the prediction $n_s \approx 0.96$ and a tensor-to-scalar ratio
$0.003 < r< 0.13$
that interpolates between the values characteristic to  quadratic~\cite{Linde:1983gd}
and to Starobinsky~\cite{Starobinsky} inflation.
In this context the smallness of the electroweak scale is connected to the smallness of the inflationary perturbations:
both arise because the underlying theory is  very weakly coupled.

In both cases, gravitational decays of the inflaton reheat the SM particles up to a temperature $10^7-10^9$~GeV.

\bigskip

The organization of the paper is the following.
In section~\ref{generic} we present our results of the effective field theory approach to dynamically induced gravity and inflation.
In section~\ref{agravity} we focus on agravity and compute inflationary parameters in this quantum gravity theory.
In section~\ref{reheating} we collect our results on reheating and on dark matter (DM) abundance of the universe.
We conclude in section~\ref{conclusions} and present technical details in Appendix~\ref{app}.

\section{Effective field theory approach}\label{generic}

In this section we present a general, model independent study of  scale-invariant single field inflation in which the
Planck scale is dynamically generated by the inflaton. The main aim of our effective field theory approach
is to derive results that are valid for all possible UV completions of gravity.
We, therefore, restrict our physical parameters such that inflation occurs within  the low-energy (sub-Planckian) limit
of gravity. 
More broadly, the hope of deriving  general implications for inflation rests on the possibility that,
if the scale symmetry is broken dynamically by a VEV induced by
{\em weakly coupled} dynamics, it leaves a {\em light} scalar,
which is the pseudo-Goldstone boson of scale invariance.
Two experimental facts support this assumption, suggesting two possibilities where an effective field theory could be adequate:
\begin{itemize}\label{uno}
\item[1)] First, the  amplitude of primordial scalar perturbations is observed to be small, $P_R=(\PRexp)\times 10^{-9}$~\cite{Planck2015}.
This suggests that  inflation occurs at a sub-Planckian energy  $E$,
where the gravitational coupling $g(E)\sim E/\bar M_{\rm Pl}$
($\bp = 2.4 \times 10^{18}$~GeV is the reduced Planck mass)
 is still small enough that no knowledge of the UV structure of quantum gravity is needed.
 We check that our effective field theory is valid in the parameter space we consider and that the results of our computations are trustable.
We will explicitly demonstrate that the results obtained in the Jordan and Einstein frames are physically equivalent.

\item[2)] Second, the smallness of Higgs boson  mass, $M_h/\bp \sim 10^{-16}$ suggests that  quantum gravity should be  weakly coupled~\cite{agravity,RGE},
such that the quantum corrections to $M_h$ are naturally small.
{\em Soft-gravity}  is the idea that the growth of the gravitational coupling $g(E)$ with energy
could be stopped by new gravitational physics at an energy $E\sim M_g$ low enough that $g(E)$ saturates at a small enough value
$g(E) \circa{<}  g(M_g)$.
Then soft-gravity can be neglected during inflation even when Einstein gravity would become non-perturbative,
extending the domain of validity of our computations.
\end{itemize}
In practice, both possibilities above amount to ignoring quantum gravity in a controllable way.

\subsection{Model-independent  dimensionless single field inflation}
Assuming no explicit mass scale in the fundamental Lagrangian,\footnote{In the full theory Lagrangian, the Higgs mass is generated via dimensional transmutation as well. We do not discuss this topic  in detail here because it depends on the exact model realisation, which is outside the scope of this section. For further details in the agravity realisation, we refer the reader to the following sections and to \cite{agravity}, where the Higgs mass was generated in such a way.} the
inflaton field $s$, singlet under the  SM gauge group, has a scalar potential consisting only of a quartic term
\begin{equation}\label{eq:Vs}
  V = \frac{1}{4} \lambda_{S}(s) s^{4},
\end{equation}
where the self-coupling $\lambda_{S}(s)$ runs due to  interactions (to be specified in
the next subsection).
The inflaton has a  non-minimal coupling to gravity $-f(s) R/2,$ where $R$ is the Ricci scalar,
parameterised by the dimensionless coupling $\xi_S$ as
\begin{equation}
  f(s) = \xi_{S} s^{2} .
  \label{eq:nonminimal}
\end{equation}
We neglect the running of $\xi_S$ in the limit of weak coupling of gravity in the Einstein frame (see Appendix~\ref{app} for more details). We assume that the SM degrees of freedom are very weakly coupled to the inflaton and do not
affect its dynamics.
For example we assume that the allowed inflaton-Higgs mixing term $s^2 |H|^2$ is negligibly small.
We will show in section~\ref{reheating} that this assumption is compatible with an acceptable reheating of the universe after inflation.

The coupling in eq.\eq{nonminimal} has the same form as the usual gravitational coupling $-{\bp^{2}} R/2$ in the Einstein-Hilbert Lagrangian.
With the assumptions made above we expect that the Planck scale and the cosmological constant must be generated by quantum corrections encoded
in the dynamics of $\lambda_{S}(s)$.
This is, indeed, possible since the running of $\lambda_{S}$ allows the scalar potential of $s$ to have a minimum at a non-zero field value.
To generate the Planck scale, the VEV $v_s$ of the inflaton field must be given by
\begin{equation}
  v_{s}^{2} = \frac{\bp^{2}}{\xi_{S}}.
  \label{eq:v:phi:Planck:mass}
\end{equation}
To compute inflationary observables we go from the Jordan frame possessing  the non-minimal coupling \rfn{eq:nonminimal} to the Einstein frame possessing the
canonical Einstein-Hilbert action of gravity with the Weyl transformation
\begin{equation}
  g_{\mu\nu}^E = \Omega(s)^{2} g_{\mu\nu}, \qquad \hbox{where}\qquad\Omega(s)^{2} = \frac{f(s)}{\bp^2}  = \frac{s^{2}}{v_{s}^{2}}.
\label{eq:conformal}
\end{equation}
The Einstein frame scalar potential is then given by
\begin{equation}
  V_E(s) = \frac{V(s)}{\Omega(s)^{4}}   = \frac{1}{4} \lambda_{S} ( s ) \frac{\bp^{4}}{\xi_{S}^{2}}.
  \label{eq:UJordan}
\end{equation}
At the minimum the value of this potential  must be (very close to) zero in order to yield the tiny positive vacuum energy density that gives the universe its current accelerated
expansion.\footnote{Notice that the `multiple point criticality' principle of~\cite{Froggatt:1995rt}
arises in the context of dynamical generation of scales because the dimensionless potential of eq.\eq{Vs} necessarily has another, unphysical, minimum
with zero cosmological constant at $s=0$.}
In our framework this requirement implies $\lambda_{S}(v_{s}) = 0$. The minimum condition on $\lambda_{S}$ is
\begin{equation}
\frac{ d\lambda_{S}}{dt}(v_{s}) = \beta_{\lambda_{S}}(v_{s})  = 0,
 \label{eq:beta:lambda:phi:at:min}
\end{equation}
where $t = \ln \bar\mu$, $\bar\mu$ is the renormalisation scale and $\beta_{\lambda_S}$ is the $\beta$ function of $\lambda_S$.
As usual, we resum log-enhanced quantum corrections by  identifying the renormalisation scale $\bar\mu$ with the inflaton field value $s$.
Moreover, in order to ensure that  $\lambda_{S}(v_{s}) = 0$ is not just a stationary point but a minimum,
we need to impose the requirement $\beta'_{\lambda_{S}} (v_{s}) \equiv  d^2\lambda_S(v_s)/dt^2 >0$.
In explicit model realisations of this scenario these requirements  imply  conditions on the model parameters.
We can Taylor expand $\lambda_S$ around the minimum $v_s$ obtaining
\begin{equation}
 \lambda_S(s) =  \frac{1}{2!} \beta_{\lambda _{S}}'(v_s) \ln^2\frac{s }{v_s} +
  \frac{1}{3!} \beta _{\lambda _{S}}''(v_s)  \ln^3\frac{s }{v_s} + \cdots ,
  \label{eq:lambdaTaylor}
\end{equation}
where we have used $\lambda_{S}(v_s)=\beta_{\lambda_{S}}(v_s)=0$.
In any model, this is a perturbative expansion that holds for small enough couplings.\footnote{Ref.~\cite{agravity} used a simpler approximation,
neglecting also the $\ln^3 s$ term. Here we investigate its impact. Of course, an extra $\ln^4 s$ term  is needed in order to stabilise the potential for $s \ll v_s$.}
Assuming weak couplings in order to get the correct small amplitude of primordial fluctuations, we will treat $\beta_{\lambda _{S}}'(v_s)$ and $\beta_{\lambda _{S}}''(v_s)$ as small constant parameters.
We will show in the next subsection that this approximation can indeed hold in the explicit model realisation.

\medskip

It is convenient to rewrite $V_E$ in terms of the canonically normalised field $s_{E}$ in the Einstein frame,
\begin{equation}
s_{E} = \sqrt{\frac{1+6 \xi_{S}}{\xi_{S} }} \bp \ln \frac{s }{v_s},
\label{eq:chi}
\end{equation}
or equivalently
\begin{equation}
  s = v_{s}  e^{\sqrt{\frac{\xi_{S}}{1+6 \xi_{S}}} \frac{s_{E}}{\bp}} .
  \label{eq:phi}
\end{equation}
Inserting (\ref{eq:lambdaTaylor}) and (\ref{eq:phi}) into (\ref{eq:UJordan}) we get
\begin{equation}
V_E(s_{E}) \simeq \frac{ \beta_{\lambda_S}'(v_s) \bp^2}{8 \xi_{S} \left(1 + 6 \xi_{S}\right)}
\left(1 +\sqrt{\frac{\xi_{S}}{1 + 6 \xi_{S}}} \frac{ \beta _{\lambda_{S}}''(v_s)}{3 \bp \beta _{\lambda_{S}}'(v_s)} s_{E}   \right)
s_{E}^2,
\label{approx}
\end{equation}
which is nothing but a quadratic potential with a cubic correction.
Such a potential is symmetric under the transformation $s_E \to - s_E$ and $\beta''_{\lambda_S}(v_s) \to -\beta''_{\lambda_S}(v_s)$, therefore by redefining the sign of $s_E$ we can always assume that $\beta''_{\lambda_S}(v_s)\ge 0$.
This potential allows for two different types of inflation:
\begin{itemize}
 \item[$-$] Negative-field  inflation, when $s_{E}$ rolls down from  negative values to zero.
 This corresponds, in the Jordan frame, to small-field inflation  ($s$ rolls down from a value $s<v_s$ to $v_s$) for $ \beta_{\lambda _{S}}''(v_s)>0$ and to
 large-field inflation  ($s$ rolls down from a value $s>v_s$ to $v_s$) if $\beta_{\lambda _{S}}''(v_s)<0$.

 \item[$+$]  Positive-field inflation, when $s_{E}$ rolls down from  positive field values to zero.
  This corresponds, in the Jordan frame, to large-field inflation for $ \beta_{\lambda _{S}}''(v_s)>0$ and to
 small-field inflation if $\beta_{\lambda _{S}}''(v_s)<0$.
\end{itemize}

\begin{figure}[t]
\begin{center}
\includegraphics[width=0.5\textwidth]{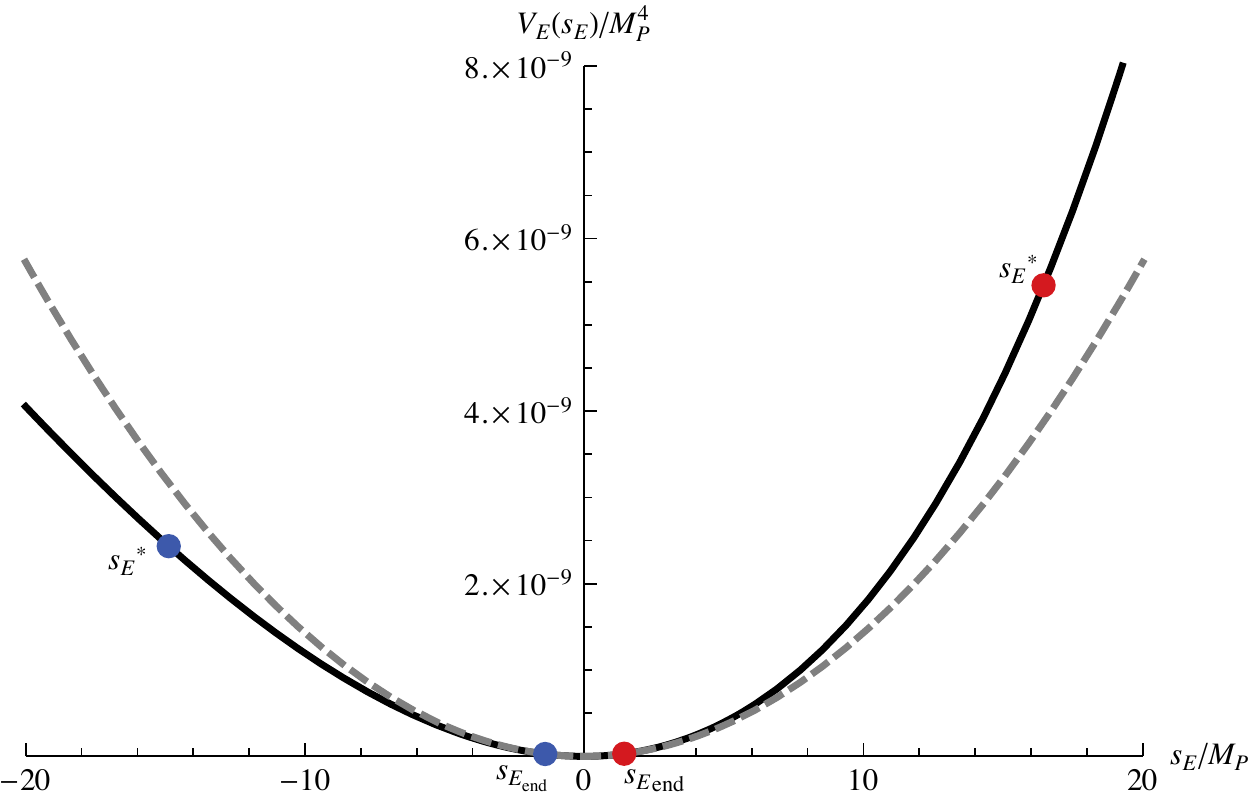}
\caption{\em The Einstein-frame inflation potential $V_E(s_{E})$ (solid) as computed in the minimal model of section~\ref{minimalmodel},
and its quadratic approximation (dashed). Blue dots show $s_{E}^{*}$ and $s_{E \rm end}$ for
 negative-field inflation and red dots for  positive-field inflation. Other parameters are specified in the text.}
\label{fig:lambda:phi:run:ind:Planck:mass}
\end{center}
\end{figure}

We present in fig.~\ref{fig:lambda:phi:run:ind:Planck:mass} an example plot of the Einstein frame potential $V_E(s_{E})$,
as computed in the minimal model presented later in section~\ref{minimalmodel}:
the potential is well approximated by the cubic potential of eq.~(\ref{approx}), with the following
values of its parameters: $\xi_{S}(v_{s}) = 300$,
$\beta'_{\lambda_{S}}(v_{s}) = 6 \times 10^{-5}$ and $\beta''_{\lambda_{S}}(v_{s}) = 9 \times 10^{-6}$.
Fig.~\ref{fig:lambda:phi:run:ind:Planck:mass} also shows the potential in quadratic  approximation (dashed parabola), which is not quite perfect.
We denote the field values corresponding to 60 $e$-folds ($s_{E}^{*}$) and to the end of inflation ($s_{E\rm end}$) with  blue dots for  negative-field inflation and with
 red dots for positive-field inflation. We follow the same colour code throughout this section.
Because of the loop-induced  cubic term in $s_{E}$,
the two inflation regimes are physically different and can be distinguished from each other experimentally. The potential in fig.~\ref{fig:lambda:phi:run:ind:Planck:mass} yields $r = 0.11$ for negative-field and $r = 0.16$ for positive-field inflation. For large $\xi_{S}$, $\beta'_{\lambda_{S}}(v_{s}) $ and $\beta''_{\lambda_{S}}(v_{s})$ (given by large couplings), higher order terms in the expansion \eqref{eq:lambdaTaylor} will become important. The cubic approximation breaks down and one has to consider numerically exact running of the couplings.

\smallskip


Under the {soft-gravity} assumption (described as point 2 at the beginning of section~\ref{generic}),
this computation holds in all its parameter space.
The model-independent approach (described as point 1) holds instead only as long as
Einstein gravity can be neglected.
It is simple  to check the validity range of the computations in the Einstein frame. The condition is
\begin{equation}
 (V_E(s_E))^{1/4} \ll \bp,
 \label{eq:Ebound}
\end{equation}
so that we can consistently ignore quantum gravity corrections.
Such a condition should be realised at least for $s_E=s_E^*$, which corresponds to the maximum potential value for the inflation computations.
Ignoring the cubic correction in eq. (\ref{approx}) we get
\begin{equation}
 \frac{1}{8} \frac{\beta_{\lambda_S}'(v_s) }{\xi_S  (6 \xi_S +1)} \ll \left( \frac{\bp}{s_E^*} \right)^2.
 \label{eq:efftheory}
\end{equation}
We considered values of $\beta_{\lambda_S}'(v_s)$, $\beta_{\lambda_S}''(v_s)$ and $\xi_S$ so that eq. \rfn{eq:Ebound} is satisfied, so that our computations are consistent and we can safely ignore quantum gravity corrections.
The consistency condition \eqref{eq:Ebound} can also be expressed in the Jordan frame as
\begin{equation}
 (V_J(s))^{1/4} \ll \sqrt\xi_S |s|,
\end{equation}
leading to the same result expressed in \rfn{eq:efftheory} after taking into account the relation between $s$ and $s_E$ (see eq.~\rfn{eq:chi}).

\medskip

To better understand how the  predictions of  negative-field inflation differ from positive-field inflation
due to the presence of the cubic term ${\beta_{\lambda _{S}}''(v_s)}/{\beta_{\lambda _{S}}'(v_s)}$,
we expand the slow-roll parameters at  first order in it.
The scalar spectral index $n_s$ and  the tensor-to-scalar ratio $r$ are given by
\begin{equation}
\begin{split}
r &\simeq \frac{8}{N} \mp \frac{32 \sqrt{2}}{9} \sqrt{\frac{\xi_S}{6 \xi_S +1}}\frac{\beta _{\lambda _{S}}''(v_s)}{\beta_{\lambda _{S}}'(v_s)}\left(\frac{1}{\sqrt{2 N}}-\frac{1}{4N^2}\right), \\
n_s &\simeq 1-\frac{r}{4} \pm \sqrt{\frac{\xi _{S}}{6 \xi _{S}+1}} \frac{\beta_{\lambda _{S}}''(v_s)}{\beta_{\lambda _{S}}'(v_s)} \frac{\sqrt{r}}{3 \sqrt{2}},
\label{eq:ns:approx}
\end{split}
\end{equation}
where $N$ denotes the number of $e$-folds, and the signs $+(-)$ should be used for the  positive-field inflation and  $-(+)$ for  negative-field inflation.
We see that (\ref{eq:ns:approx}) predicts somewhat different behaviour of  $n_s$ and $r$ for the positive-field and negative-field inflation scenarios.

The approximation \eqref{eq:ns:approx} breaks down if $\xi_{S}$ and other couplings are large. In that case the deviation of $r$ from quadratic inflation can be large too, as seen in the minimal model realisation presented in subsection \ref{minimalmodel}.

\medskip

 In conclusion, the observed small value of $P_R$ favours a small inflaton self-coupling. If other couplings are small as well, then the Einstein-frame inflaton potential is well approximated by a quadratic potential. If other couplings are large, the deviation of $r$ from quadratic inflation can be strong, as shown in fig.~\ref{fig:r:vs:ns:ind:Planck:mass}.
In  section~\ref{agravity} we will show that in agravity~\cite{agravity} --- a concrete UV completion of gravity --- dimensionless inflation can give a significantly smaller value of $r$ if all couplings are small.
Basically this will arise because agravity realises the soft-gravity scenario by adding to the Lagrangian
dimensionless terms of $R^2$ form (as in Starobinsky inflation), leading to extra light scalars.

\subsection{The minimal model for dimensionless single field inflation}\label{minimalmodel}

In this subsection we present the  minimal model that dynamically reproduces all features of dimensionless single field inflation considered in the previous subsection.
Besides the inflaton $s$, the minimal model contains another real scalar $\sigma$ and a Majorana fermion $\psi$.
This is the minimal field content that is needed to achieve condition \eqref{eq:beta:lambda:phi:at:min} dynamically.
Indeed, the portal coupling of the inflaton with the extra scalar is needed to trigger dimensional transmutation while the extra
fermion is needed to be able to tune the minimum of the potential according to the multiple point criticality principle.
The latter is possible since the scalar and fermion couplings contribute to the running of the inflaton self-coupling with opposite signs,
as is apparent from the RGEs presented in appendix~\ref{app}. This fact has also been used to achieve the multiple point criticality in
Higgs inflation~\cite{Hamada:2014xka}.

Thus the Jordan frame Lagrangian of the minimal model is
\begin{align}
  \sqrt{- g^{J}} \mathscr{L}^{J} &= \sqrt{- g^{J}} \left[ \mathscr{L}_{\rm SM} - \frac{\xi_{S}}{2} s^{2} R
  + \frac{(\partial s)^{2}}{2}  + \frac{(\partial \sigma)^{2}}{2}  + \frac{i}{2} \bar{\psi}^{c} \slashed{D} \psi + \mathscr{L}_{Y} - V \right],
  \\
  \mathscr{L}_{Y} &= \frac{1}{2} y_{S} s \bar{\psi}^{c} \psi + \frac{1}{2} y_{\sigma} \sigma \bar{\psi}^{c} \psi,
  \\
   V &= \frac{1}{4} \lambda_{S} s^{4} + \frac{1}{4} \lambda_{S\sigma} s^{2} \sigma^{2} + \frac{1}{4} \lambda_{\sigma} \sigma^{4},
   \label{eq:Jordan:Lagrangian}
\end{align}
where we neglected the couplings to the SM fields, as suggested by the hierarchy problem.
In the Einstein frame, the Lagrangian reads
\begin{align}
  \sqrt{- g^{E}} \mathscr{L}^{E}  &= \sqrt{- g^{E}} \left[ \frac{ \mathscr{L}_{\rm SM} }{ \Omega(s)^{4} } - \frac{1}{2} \bp^{2} R
  + \frac{ (\partial s_{E})^{2}}{2} + \frac{(\partial \sigma_E)^{2}}{2}  + \frac{i}{2} \bar{\psi}_E^{c} \slashed{D} \psi_E + \mathscr{L}_{Y_E} - V_E + \cdots \right],
\label{eq:Einstein:Lagrangian}
  \\
  \mathscr{L}_{Y_E} &= \frac{1}{2} y_{S} v_{s} \bar{\psi}_E^{c} \psi_E + \frac{1}{2} y_{\sigma} \sigma_E \bar{\psi}_E^{c} \psi_E
  \equiv \frac{1}{2} m_{\psi} \bar{\psi}_E^{c} \psi_E + \frac{1}{2} y_{\sigma} \sigma_E \bar{\psi}_E^{c} \psi_E,
  \label{eq:Einstein:Yukawa}
  \\
   V_E &= \frac{1}{4} \lambda_{S} v_{s}^{4} + \frac{1}{4} \lambda_{S\sigma} v_{s}^{2} \sigma_E^{2} + \frac{1}{4} \lambda_{\sigma} \sigma_E^{4}
   \equiv \Lambda + \frac{1}{2} m_{\sigma}^{2} \sigma_E^{2} + \frac{1}{4} \lambda_{\sigma} \sigma_E^{4},
   \label{eq:Einstein:potential}
\end{align}
where, in order to have canonical kinetic terms, the Einstein-frame scalar and fermion fields are defined as
\begin{equation}
  \sigma_{E} = \frac{\sigma}{\Omega(s)}, \quad   \psi_{E} = \frac{\psi}{\Omega(s)^{\frac{3}{2}}},
  \label{eq:Einstein:fields}
\end{equation}
whereas gauge vectors are invariant under the transformation.
It can be shown that the derivative of the denominator in \eqref{eq:Einstein:fields} cancels out in the fermion kinetic term because of the spin connection contribution in $\slashed{D}$~\cite{Hayashi:1976uz,Maeda,Watanabe}, whereas for scalars (with the exception of the canonically normalised inflaton field $s_{E}$) it induces a derivative interaction. For simplicity we omit these details in the above Lagrangian, which are essential for reheating the universe
after inflation and will be discussed in section~\ref{reheating}. Below, we work in the Einstein frame and omit indices (except for $s_{E}$).

Note that in the  scalar potential and in the Yukawa terms of the high scale inflationary physics the scale transformation is equivalent to the substitution $s \to v_{s}$ and, therefore, in the Einstein frame the fermion $\psi$ and the scalar $\sigma$ do not have couplings to the inflaton at tree-level. The Jordan frame self-coupling term of the inflaton becomes the cosmological constant $\Lambda$ in the Einstein frame potential \eqref{eq:Einstein:potential} (equivalent to \eqref{eq:UJordan}). The Yukawa and quartic portal terms of the inflaton become mass terms, giving
 the mass of the $\sigma$ field in the Einstein frame by
\begin{equation}
  m_{\sigma}^2  = \frac{1}{2} \lambda_{S\sigma}(v_s) \frac{\bp^{2}}{\xi_{S}},
\end{equation}
and the mass of the fermion $\psi$ in the Einstein frame by
\begin{equation}
  m_{\psi}  = y_{S}(v_s) \frac{\bp}{\sqrt{\xi_{S}}}.
\end{equation}

 The scalar potential depends on the inflaton field only due to the running of the scalar couplings.

The renormalisation group equations (RGEs) of the model in the weakly coupled gravity limit are computed in appendix~\ref{app}. In the Jordan frame, gravity does not contribute to the running of the couplings at the one-loop level \cite{Herranen:2014cua}. The transformation to the Einstein frame mixes the gravitational and scalar degrees of freedom such that in the Einstein frame the matter RGEs
get contributions from gravity, that we neglect.  We can explicitly verify the equivalence of the frames  only up to these neglected effects, as discussed in appendix~\ref{app} around eq.~\eqref{eq:good:limit}.


\medskip

We suppose that inflation  takes place along the $s$ field direction (that is, $\sigma=0$). We will see later that such an assumption is self-consistent: since the scalar $\sigma$ will turn out to be heavier than the inflaton, it does not take part in inflation.
As discussed earlier, we need to realise $\lambda_{S}(v_{s}) = \beta_{\lambda_{S}}(v_{s}) = 0$.
Imposing $\lambda_S(v_s)=0$, the second condition becomes (see eq.\eq{blambda}),
\begin{equation}
  16 \pi^2 \beta_{\lambda_{S}}(v_{s}) = \frac{1}{2} \lambda_{S\sigma}^2 - 4 y_{S}^{4} = 0,
 \label{eq:y:lambda:phieta:at:min}
\end{equation}
Moreover, in order to ensure that $v_s$ is not just a stationary point but a minimum, we need to impose that $\lambda_{S}^{\prime\prime} (v_{s}) = \beta_{\lambda_{S}}^{\prime}(v_{s}) > 0$.
At the minimum
\begin{equation}
\begin{split}
  \beta_{\lambda_{S}}^{\prime}(v_{s}) &= \frac{1}{16 \pi^{2}} \left[ 8 y_{S} (\lambda_{S} -2 y_{S}^2) \beta_{y_{S}}
   + 4 (9 \lambda_{S} + y_{S}^2) \beta_{\lambda_{S}} + \lambda_{S\sigma} \beta_{\lambda_{S\sigma}} \right] \\
  &= \frac{ \lambda_{S\sigma}^{2} \left[ 6 \lambda_{\sigma} + (4 - \sqrt{2}) \lambda_{S\sigma} - (4 + 6 \sqrt{2}) y_{\sigma}^{2} \right] }{256 \pi^{4}},
\end{split}
\label{eq:lambda:phi:ii}
\end{equation}
where we have used $\lambda_{S}(v_{s}) = 0$ and the relations \eqref{eq:v:phi:Planck:mass} and \eqref{eq:y:lambda:phieta:at:min}.


\begin{figure*}[t]
\begin{center}
\includegraphics[width=0.47\textwidth]{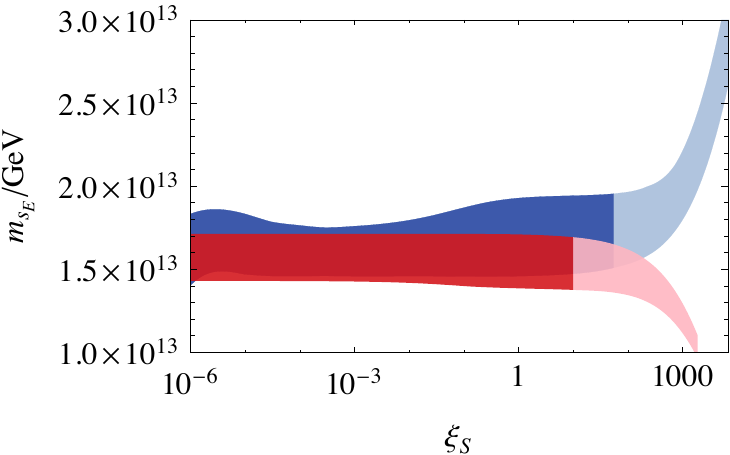}\qquad
\includegraphics[width=0.43\textwidth]{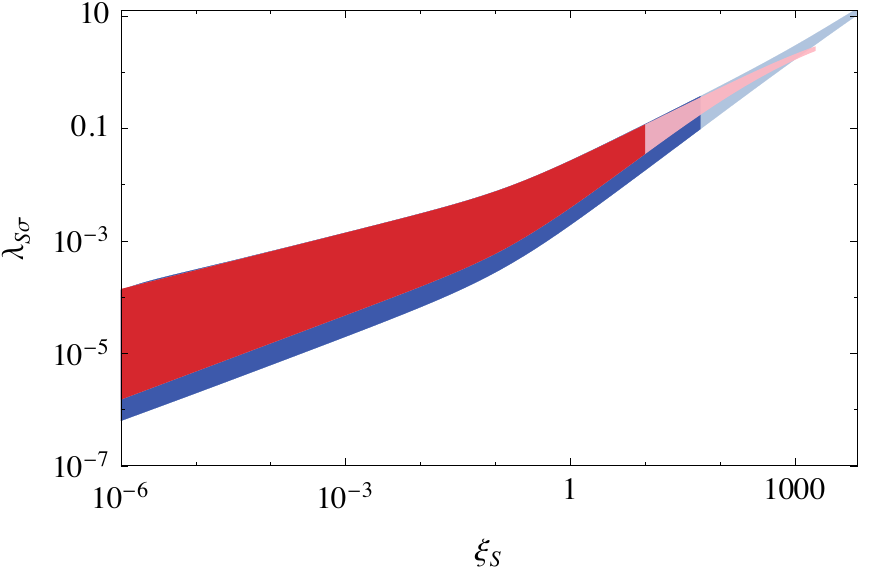}
\caption{\em Acceptable range of parameters in the minimal dimensionless inflation model.
{\bf Left}: inflaton mass $m_{s_{E}}$ as a function of $\xi_{S}$. {\bf Right}: the inflaton-$\sigma$ coupling
 $\lambda_{S\sigma}$ as a function of $\xi_{S}$. Blue means negative-field and red means positive-field inflation. Lighter colours mark the regions where gravity cannot be ignored in the Einstein frame.}
\label{fig:lambdaphisigmavsxi}
\end{center}
\end{figure*}

\begin{figure}[t]
\begin{center}
\includegraphics[width=0.5\textwidth]{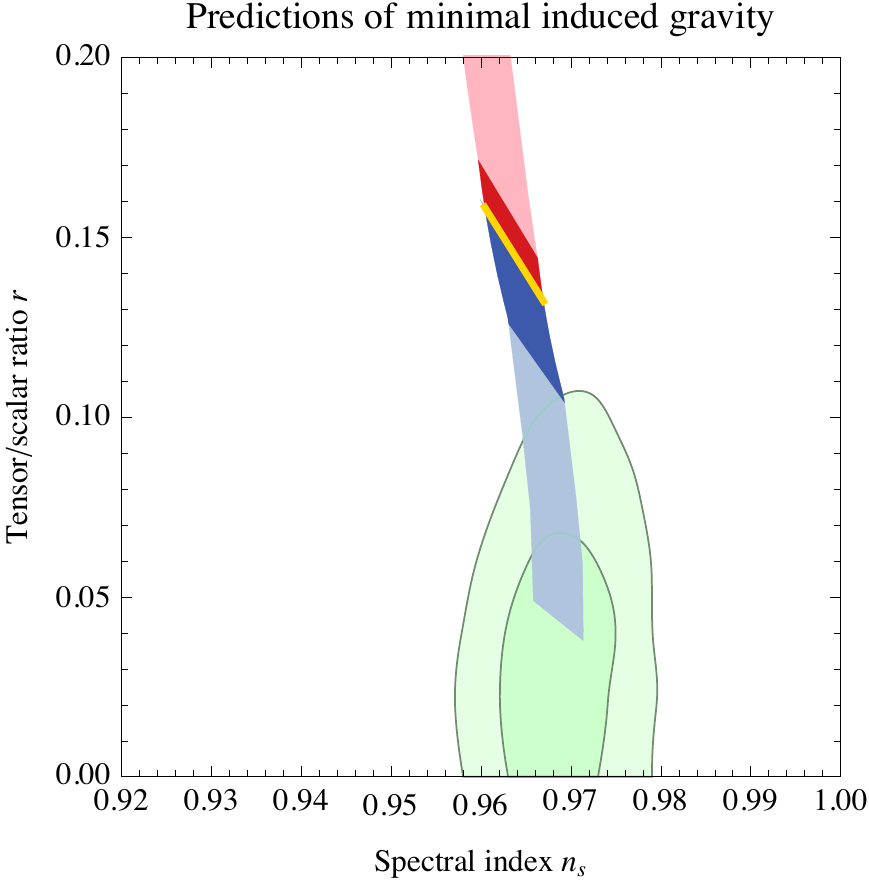}
\caption{\em Predictions of dimensionless single field inflation in the $(n_s,r)$ plane for $50<N<60$ $e$-folds
assuming that gravity is described by the Einstein action, and that it gives small quantum corrections.
The blue region shows the prediction of negative-field inflation and red shows positive-field inflation around the quadratic potential (the yellow line). Light blue and light red mark the regions where gravity cannot be ignored in the Einstein frame.
The light green contours are the 1,2$\sigma$ best fit regions from {\sc Planck, BICEP2/Keck}~\cite{BICEP2/Keck:2015tva,Ade:2015fwj,Planck2015}.}
\label{fig:r:vs:ns:ind:Planck:mass}
\end{center}
\end{figure}

Physically, it means that the inflaton mass in the Einstein frame,
\begin{equation}
  m_{s_{E}}^2 = \frac{ \beta_{\lambda _s}'(v_s) \bp^2}{4 \xi _{S} \left(1 + 6 \xi _{s}\right)},
\end{equation}
has to be positive. Since $m_\sigma$ arises at tree-level, while $m_{s_E}$ arises at loop level,
the $\sigma $ field is typically heavier than the inflaton field and remains frozen at its minimum  $\sigma=0$ during inflation.
We thereby have realised the single field  scenario discussed in the previous subsection. The presence of a non-vanishing Yukawa coupling $y_{S}$ is needed in order to realise dynamical generation of the Planck scale
with a vanishing cosmological constant; this implies that $m_{\psi}$
must be larger than the inflaton mass $m_{s_{E}}$. If the model has more than one fermion,
some of them can be lighter than $s_E$.

\medskip

Taking into account all constraints, we determine the allowed parameter region for the minimal model.
In fig.~\ref{fig:lambdaphisigmavsxi} we plot the inflaton mass as a function of $\xi_S$
 for both negative-field and positive-field inflation (the two regions are partially overlapping) for $50<N< 60$~\cite{N}.
The non-minimal coupling cannot be arbitrarily small, because that would mean $v_{s} \gg \bp$, implying  trans-Planckian masses for $\sigma$ and $\psi$. Trans-Planckian masses are avoided for $\xi_S \circa{>} 10^{-6}$. An upper bound on the non-minimal coupling comes from the requirement that gravity corrections to couplings in the Einstein frame can be neglected (this is true if conditions \eqref{eq:good:limit} hold, see Appendix \ref{app}). The bound is $\xi_{S} \circa{<} 10(50)$, for positive-field (negative-field) inflation. In the region where gravity corrections cannot be neglected in the Einstein frame, the range of $\xi_{S}$ can be extended (shown with lighter colours) up to $\xi_{S} \simeq 1900$ for positive-field and $\xi_{S} \simeq 6900$ for negative-field inflation.\footnote{Notice that a large $\xi$-coupling does not necessarily spoil perturbative unitarity if the VEV of the corresponding scalar field is large \cite{Bezrukov:2010jz,Giudice:2010ka}.
}  In this case the upper bound on $\xi_{S}$ arises from perturbativity of running couplings: $\lambda_{\sigma} \circa{<} 2 \pi/3$ and $\lambda_{S\sigma} \circa{<} 4 \pi$. The range of $\lambda_{S\sigma}(v_{s})$ in the right panel of fig.~\ref{fig:lambdaphisigmavsxi} is determined by the normalisation of the spectrum $P_R = {V_E(s^{*})}/{24 \pi^{2}{\bp^{4}}  \epsilon (s^{*})}$.
The running of $\lambda_{S\sigma}$ depends on the self-coupling $\lambda_{\sigma}$. If $\lambda_{\sigma}(v_{s})$ is relatively small, then $\lambda_{S\sigma}(v_{s})$ has to be larger at the potential minimum. If $\lambda_{\sigma}(v_{s})$ is large, then $\lambda_{S\sigma}$ runs faster to the required value at $s^{*}$ and can be smaller in the minimum. Variations of $\lambda_{S\sigma}$ and $\lambda_{\sigma}$ determine the range of the inflaton mass in the left panel of fig.~\ref{fig:lambdaphisigmavsxi} as well.

The inflaton potential in  fig.~\ref{fig:lambda:phi:run:ind:Planck:mass} corresponds to the model parameters
 $\xi_{S}(v_{s}) = 300$, $\lambda_{S\sigma}(v_{s}) = 0.725$,  $\lambda_{\sigma}(v_{s}) = \pi/12$, $y_{S}(v_{s}) = 0.253$,
 $y_{\sigma}(v_{s}) = 0$ and $m_{s_{E}} = 1.28\times 10^{13}$~GeV that are in the physical range, verifying our model independent results in the previous subsection.\footnote{We choose $y_{\sigma}(v_{s}) = 0$ for simplicity. If $y_{\sigma} \neq 0$, the predictions for inflation do not change, since its negative influence on the running of $\lambda_{S\sigma}$ must be countered by a larger value of $\lambda_{\sigma}$ in order to get the correct value for $P_{R}$.}  The large values of the couplings are needed to get a significant deviation from quadratic inflation.

The slow-roll parameters in this model are given by
\begin{equation}
  \epsilon
  = \left[ \frac{ \lambda_{S}^{\prime} (s) }{ \lambda_{S} ( s ) } \right]^{2}
 \frac{\xi_{S} s^{2}}{2 (1 + 6 \xi_{S})},
  \qquad
  \eta
  = \frac{ s \xi_{S} [ \lambda_{S}^{\prime} ( s ) + s \lambda_{S}^{\prime\prime} (s ) ] }{\lambda_{S} ( s ) (1 + 6 \xi_{S}) },
\end{equation}
in terms of which the inflationary parameters are given by
\begin{equation}
n_{s} = 1 - 6 \epsilon(s^{*}) + 2 \eta(s^{*}), \qquad r = 16 \epsilon(s^{*}).
\end{equation}
The predictions of dimensionless single field inflation for $r$ as a function of $n_s$ are presented in fig.~\ref{fig:r:vs:ns:ind:Planck:mass} for $50<N< 60$~\cite{N}. To compare our predictions with experimental results we plot in the same figure also the contours of
  the 1,2 $\sigma$ best-fit regions from the official combination of the BICEP2/Keck Array/Planck \cite{BICEP2/Keck:2015tva,Ade:2015fwj,Planck2015}.

The yellow line represents the
quadratic approximation obtained in the limit ${\beta_{\lambda _{S}}''(v_s)}=0$ (see eq. (\ref{approx})). The blue region
shows the allowed parameter space for  negative-field inflation and the red region for the positive-field inflation. In the region with darker colours around the yellow line, the conditions \eqref{eq:good:limit} hold and gravity corrections can be neglected in the Einstein frame. In this region, $\lambda_\sigma$ and $\lambda_{S\sigma}$ are small and the inflaton potential is well approximated by \eqref{approx}. The predictions for the inflationary parameters  $n_{s}$ and $r$ roughly coincide with the model-independent predictions.
 In the light red and light blue regions, the conditions \eqref{eq:good:limit} do not hold, but due to the equivalence of the frames the gravitational corrections in the Einstein frame must arise from scalar loops in the Jordan frame. The potential is not close to the cubic \eqref{approx} any more. We see that for large couplings taking into account exact numerical solutions for RGE running can induce a large correction in $r$. For the number of $e$-folds $N = 60$, the lowest possible value of the tensor-to-scalar ratio is $r = 0.04$.

\section{Inflation in agravity}\label{agravity}
In this section we reconsider  inflation  within agravity~\cite{agravity}: a renormalizable extension of Einstein gravity, obtained by adding all
dimensionless couplings which are anyhow generated by quantum corrections,
and removing any massive parameter such that power divergences must vanish.
The action has the generic structure
\beq \label{eq:Ladim}
\mathscr{S} = \int d^4x \sqrt{|\det g|} \,\bigg[ \Lag_{\rm matter}-\sum_i \xi_i \frac{\varphi_i^2}{2} R
+\frac{R^2}{6f_0^2} + \frac{\frac13 R^2 -  R_{\mu\nu}^2}{f_2^2}
\bigg].
\eeq
The gravitational kinetic terms suppressed by the dimensionless constants $f_0$ and $f_2$
contain four derivatives:
thereby the graviton contains a massive spin-2 ghost component~\cite{Stelle}, which is possibly problematic for energies above its mass $M_2= |f_2| {\bar M}_{\rm Pl}/\sqrt2$:
we do not address the issue of finding a sensible interpretation for it  (see \cite{Tomboulis:1977jk} for some attempts). In this section, as mentioned in the introduction,  we do not adopt the effective field theory approach of section \ref{generic} and eq. (\ref{eq:Ladim}) is assumed to be the full action, like in ref. \cite{agravity}.
As a curiosity, we notice that the classical gravitational equations of motion,
in a theory with neither matter nor cosmological constant,
have inflationary solutions with arbitrary Hubble constant. 
\smallskip

Of course,
matter must be present in a realistic theory: a generic
$\Lag_{\rm matter}$ can be written in terms of real scalars $\varphi$, Weyl fermions $\psi$ and vectors $A_\mu$ with
gauge, Yukawa and quartic couplings $g$, $y$ and $\lambda$.
Furthermore, the scalars $\varphi_i$ can have dimensionless $\xi_i$ couplings to gravity.
Once that scalars dynamically get a vacuum expectation value generating the Planck mass as $\sum_i\xi_i \varphi_i^2 = \bp^2$,
agravity realises the scenario of soft-gravity: the graviton splits into the usual graviton, a
massive spin-2 ghost-like graviton  and a scalar; their masses $M_2$ an $M_0$ represent the energy
scale at which gravity softens, becoming described by the dimensionless couplings $f_2$ and $f_0$.


The theory is renormalizable, and
quantum corrections enhanced by large logarithms are taken into account, as usual,
by substituting the couplings with running couplings
(RGE equations have been computed in~\cite{agravity}),
renormalised at an energy  comparable to the energy or field value of the process under consideration.

\bigskip

\subsection{Agravity in the Einstein frame}
We want to employ the results in the literature that give the inflationary predictions of multifield Einstein gravity models.
Then, we need to recast the  agravity action of eq.~(\ref{eq:Ladim}) in Einstein form.
We here use a compact notation, leaving implicit the sums over the scalars $\varphi_i$, which, in a realistic theory,
include at least the Planckion $s$, the physical Higgs $h$ and the other components of the Higgs doublet $H$.

\smallskip

We start adding to the generic agravity Lagrangian the vanishing term $ - {(R+3f_0^2 \chi/2)^2}/{6f_0^2}$, where $\chi$ is an auxiliary field with no kinetic term.
Such new term is designed to  cancel  $R^2/6f_0^2$, leaving
\beq \Lag=
\sqrt{|\det g|} \,\bigg[ \Lag_{\rm matter} +\frac{\frac13 R^2 -  R_{\mu\nu}^2}{f_2^2} -\frac{f}{2} R -\frac{3f_0^2}{8} \chi^2\bigg],
\eeq
where $f = \chi +\xi \varphi^2$
and\footnote{If one makes  the sum over the scalars $\varphi_i$ explicit, one should read $\xi \varphi^2$ as $\sum_i\xi_i \varphi_i^2$, the kinetic term $(D_\mu \varphi)^2$ as $\sum_i D_\mu \varphi_iD_\mu \varphi_i$ and so on.}
\beq \Lag_{\rm matter} = \frac{(D_\mu \varphi)^2}{2} -\frac14 F_{\mu\nu}^2 +\bar\psi i \slashed{D} \psi
+(y \,\varphi\psi\psi+\hbox{h.c.}) - V(\varphi).\eeq
Here $y$ denotes a set of Yukawa couplings and $V$ is a general quartic potential.
Next, we transform the $-fR/2$ term into the canonical Einstein term $-\bar M_{\rm Pl}^2 R_E/2$
by performing a rescaling of the metric,
\beq g^E_{\mu\nu}= g_{\mu\nu}\times  f/\bar M_{\rm Pl}^2 .\eeq
In the limit of constant $f$ (global scale transformation) our dimensionless action is invariant
provided that the scalars $\varphi$, the fermions $\psi$ and the vectors $A_\mu$ are also rescaled as:
\beq \varphi_E=  \varphi \times (\bp^2/f)^{1/2} ,\qquad
\psi_E = \psi \times (\bp^2/f)^{3/4},\qquad
A_{\mu E} = A_\mu . \label{Einstein-Jordan}
\eeq
However, we need to consider a non-constant $f$ and perform a local scale transformation, under which
all dimensionless terms without derivatives remain (trivially) invariant.
Furthermore, various kinetic terms happen to be also (non-trivially) invariant:
this is the case for the fermion kinetic terms~\cite{Hayashi:1976uz,Maeda,Watanabe}, the vector kinetic terms and the graviton kinetic term proportional to $1/f_2^2$.
The scalar kinetic terms are not invariant (away from the special conformal value $\xi=-1/6$);
thereby we keep using $\varphi$ in addition to $\varphi_E$ for the scalars.
Then the Einstein-frame Lagrangian is:
\beq \label{eq:Einstein}
\Lag =\sqrt{\det g_E} \bigg[  \frac{\frac13 R_E^2 -  R_{E\mu\nu}^2}{f_2^2}
 -\frac14 F_{E\mu\nu}^2 +\bar\psi_E i \slashed{D} \psi_E +
(y \varphi_E \psi_E\psi_E+\hbox{h.c.})- \frac{\bar M_{\rm Pl}^2}{2} R_E +  \Lag_\varphi - V_E  \bigg],
\eeq
{where}
\beq \Lag_\varphi =
 \bar M_{\rm Pl}^2 \bigg[ \frac{(D_\mu \varphi)^2}{2f}
 +  \frac{3 (\partial_\mu f)^2}{4 f^2}\bigg]  ,\qquad
V_E = \frac{\bar M_{\rm Pl}^4}{f^2}\bigg[{V(\varphi)}+   \frac{3f_0^2}{8} \chi^2\bigg] .  \eeq
A  kinetic
term for $f$ has been generated~\cite{fkin}, such that $f$ becomes an extra scalar, with no gauge charge.\footnote{The scalar kinetic term is conformally
invariant for $\xi_\varphi=-1/6$; this manifests as cancellations in the scalar kinetic terms.}
The kinetic metric in scalar field space has constant {\em negative} curvature $- N_\varphi(N_\varphi+1)\bar M_{\rm Pl}^2/6$, where $N_\varphi$ is the total number of scalars $\varphi$, and can be conveniently put
in conformal form by redefining
$z=\sqrt{6 f}$,
such that our final Lagrangian is
\beq \label{eq:LKin}
 \Lag_\varphi =   \frac{6\bp^2}{z^2}
 \frac{(D_\mu \varphi)^2 + (\partial_\mu z)^2}{2}
\eeq
and
  \beq \label{eq:VE}
V_E(z,\varphi)= \frac{36\bp^4}{z^4}
\bigg[{V(\varphi)}+   \frac{3f_0^2}{8} \bigg(\frac{z^2}{6} - \xi_\varphi  \varphi^2\bigg)^2\bigg] .\eeq

We anticipate here a non-trivial peculiarity of the Einstein-frame Lagrangian, best seen by considering
the case of a single `Planckion' scalar field $s$, such that $f=\xi_S s^2$: by using the first equation in (\ref{Einstein-Jordan}) we obtain that $s_E = \bp/\sqrt{\xi_S}$ is a constant
i.e.\ its quartic becomes a cosmological constant and its Yukawa couplings become fermion mass terms.
How can this be equivalent to the Jordan frame Lagrangian where $s$ has quartic and Yukawa interactions?
The point is that $s$, being the pseudo-Goldstone boson of spontaneously broken approximate scale invariance
(the explicit breaking of scale invariance coming from the quantum running of the coupling constants
is small because we are assuming perturbative couplings),
couples to  the divergence of the dilatation current $\mathscr{D}_\mu$, $\partial_\mu \mathscr{D}_\mu$, that vanishes at tree-level because
we consider special dimensionless theories.\footnote{The explicit verification that the Jordan frame couplings of $s$ vanish on-shell needs manipulations similar to the ones used to verify the analogous property of the couplings of a Goldstone boson of a U(1) global symmetry, when a Dirac fermion mass term $\bar\Psi \Psi$ is re-expressed as derivatives acting within a chiral current $\bar \Psi \gamma_\mu\gamma_5\Psi$.}


\subsubsection*{Mass eigenstates}
We compute here the mass eigenstates formed by the scalars $\phi=\{h,s,z\}$ at the minimum of the potential,
where  the scalars kinetic terms of eq.\eq{LKin} become canonical.
Indeed, minimisation with respect to $z$ leads to $z^2= 6\bp^2 + 16 V/f_0^2\bp^2$.
Minimisation with respect to $s$  gives  ${\partial V}/{\partial s} - {4\xi_S sV}/{\bp^2}=0$,
that should be solved by $\bp^2=\sum \xi_i \varphi_i^2 \simeq \xi_S s^2$.
The measured value of the cosmological constant implies a negligible value of $V$ at the minimum, simplifying the above equations.
Minimisation with respect to $h$  then leads to a
negligible vacuum expectation value.  On the other hand gauge invariance implies that $h$ should appear at least quadratically in $V$; therefore expanding $h$ around its VEV necessarily produces at least one power of this negligible VEV, which implies that the Higgs negligibly mixes with $s$ and $z$.
The mass matrix for the  fields $s$ and $z$  around the minimum is given by the second derivatives of $V_E$:
\beq \label{eq:massmatrix}
M_s^2
\begin{pmatrix}
1&0\cr 0 & 0
\end{pmatrix}+ \frac{f_0^2\bar M_{\rm Pl}^2}{2}
\begin{pmatrix}
6\xi_S&\sqrt{6\xi_S}\cr \sqrt{6\xi_S} & 1
\end{pmatrix} .
\eeq
The first term alone would give a Planckion with mass $M_s^2 =  \partial^2 V/\partial s^2$.
The second term alone would give a spin-0 graviton with mass
 $M_0^2=\frac{1}{2} f_0^2\bar M_{\rm Pl}^2(1+6\xi_S)$.
Taking into account both terms, the mass eigenvalues are
\beq\label{eq:mm}
M_\pm^2 =  \frac{M_s^2 + M_0^2}{2} \pm \frac12  \sqrt{(M_s^2+M_0^2)^2 - 4 \frac{M_s^2 M_0^2}{1+6\xi_S}}.
\eeq


\subsection{Computing multifield inflationary predictions} \label{multifield inflationary predictions}
The classical equations of motion for the Einstein-frame scalar fields  $\phi = \{z,s,h\}$ during inflation in slow-roll approximation are
\beq
\frac{d\phi}{dN} = - \frac{z^2}{6V_E}  \frac{\partial V_E}{\partial \phi},\qquad
\eeq
having defined the number of $e$-folds $N$ as $dN = H\,dt$.
The spin-2 massive graviton does not affect such classical equations of motion, and we assume that it can be neglected even at the quantum level.
The quantum predictions for inflation can now be computed by using the previous literature on multifield inflation~\cite{info};  they can be expressed in terms of
the number of $e$-folds starting from a generic initial point, $N(h,s,z)$:
\begin{itemize}
\item The power-spectrum of scalar fluctuations is given by
\be P_R(k)=\left(\frac{H}{2\pi}\right)^2  \frac{z^2}{6 \bp^2} (\nabla N)^2 \ee
with $H$ computed at horizon exit $k=aH$ and
\be (\nabla F)^2 \equiv  \left(\frac{\partial F}{\partial z} \right)^2+ \left(\frac{\partial F}{\partial s} \right)^2+ \left(\frac{\partial F}{\partial h} \right)^2.  \ee

\item  The spectral index $n_s$ of scalar perturbations is given by
\bea n_s&\equiv& 1+\frac{d\ln P_R}{d\ln k}=
\frac{1}{6 z^2V_E^2 (\nabla N)^2}\left\{6V_E^2\bigg(z^2 (\nabla N)^2-12\right)-z^4 (\nabla N)^2
(\nabla V_E)^2  \nonumber\\
&&\left.+2z^3V_E\left[\left(\frac{\partial N}{\partial h} \right)^2 \left(z\frac{\partial^2 V_{E}}{\partial h^2}- \frac{\partial V_{E}}{\partial z}\right)+\left(\frac{\partial N}{\partial z} \right)^2 \left(z\frac{\partial^2 V_{E}}{\partial z^2}+ \frac{\partial V_{E}}{\partial z}\right)\right.\right.\nonumber \\ &&
\left.\left.+2\frac{\partial N}{\partial h}\left( \frac{\partial N}{\partial z}\left(z\frac{\partial^2 V_{E}}{\partial z\partial h}+ \frac{\partial V_{E}}{\partial h}\right) +z \frac{\partial N}{\partial s} \frac{\partial^2 V_E}{\partial s \partial h}\right) +2\frac{\partial N}{\partial z}\frac{\partial N}{\partial s}\left(z\frac{\partial^2 V_{E}}{\partial s\partial z}+\frac{\partial V_{E}}{\partial s}\right)\right. \right. \nonumber \\ &&+
 \left(\frac{\partial N}{\partial s} \right)^2 \left(z\frac{\partial^2 V_{E}}{\partial s^2}- \frac{\partial V_{E}}{\partial z}\right) \bigg] \bigg\}. \eea

\item The tensor power spectrum is given by
$P_t(k) = ({2}/\bar M_{\rm Pl}^2) ({H}/{2\pi})^2$.
Equivalently, the tensor-to-scalar ratio is  given by
\be r  \equiv \frac{4 P_t}{P_R} =  \frac{48}{z^2( \nabla N)^2}.\ee

\end{itemize}
The measured values at $k= 0.002\, \mbox{Mpc}^{-1}$ are
$ P_R(k) = (\PRexp) \times 10^{-9}$~\cite{Planck2015},
$n_s=   0.965 \pm 0.006$~\cite{Ade:2013uln,Planck2015} and $r=0.06\pm0.04$ (according to~\cite{BICEP2/Keck:2015tva})
or $r=0\pm0.04 $ (according to~\cite{Planck2015}).

\begin{figure}[t]
\begin{center}
$$\includegraphics[width=0.45\textwidth]{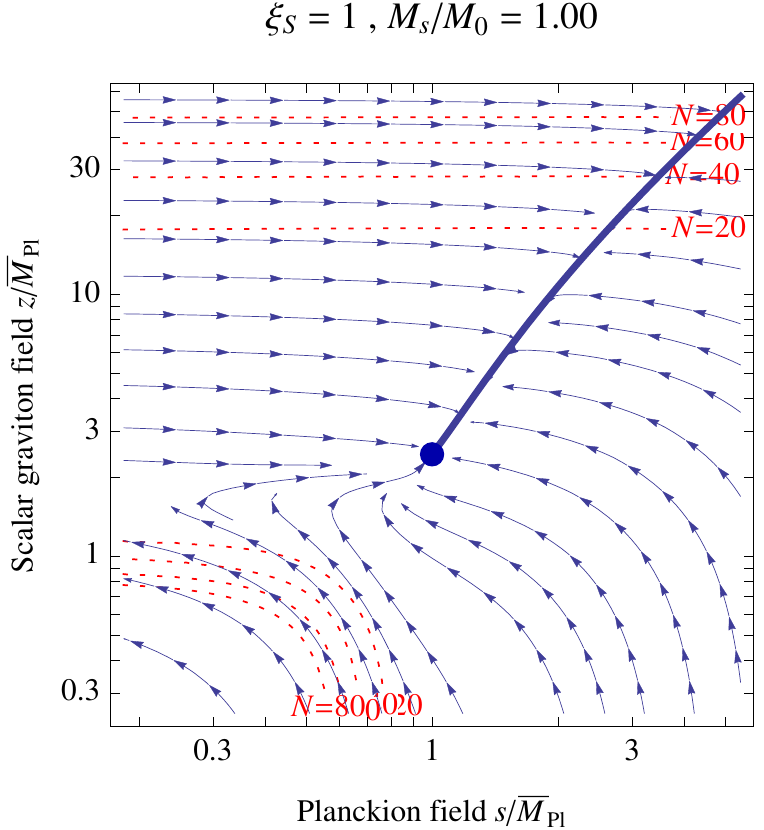}\qquad
\raisebox{0.5cm}{\includegraphics[width=0.45\textwidth]{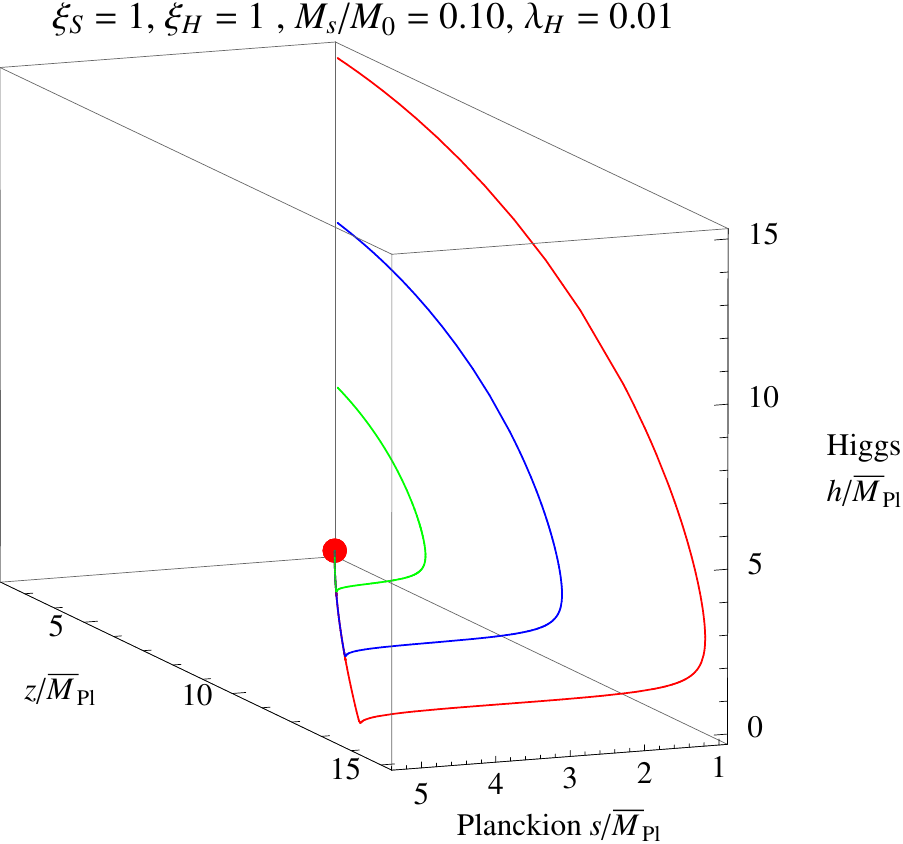}}
$$
\caption{\em {\bf Left:} sample of classical field evolution in the $(s,z)$ plane
with $h=0$  for $\xi_S=1$ and $M_s=M_0$.
All trajectories starting from different initial values
converge towards a unique inflationary attractor (thick curve) that ends at the minimum
(thick dot).
The red dashed contour-lines show the number $N$ of $e$-foldings.
{\bf Right}: sample of classical field evolution in the $(h,s,z)$ space, showing that
an attractor solution with negligible $h$ is reached
even starting from $h\gg s,z$.
}
\label{infatt}
\end{center}
\end{figure}

\subsection{Inflationary predictions}
In general, predictions of multifield inflation depend on the inflationary trajectory reducing the predictive power.
However, our potential $V_E(h,s,z)$ has a peculiar structure, such that all classical trajectories converge towards a unique attractor solution even when scalar masses are comparable at the minimum (examples are shown in fig.~\ref{infatt}).
This presumably happens because we are considering dimensionless dynamics,
such that the derivatives of the potential are hierarchical almost everywhere in field space.
We find that slow-roll inflation starts only when such attractor is reached.
In order to understand our results, it is useful to first consider three relevant extreme limits:
\begin{enumerate}
\item {\bf Planckion inflation}.
If $M_0\gg M_s$ (obtained when the agravity coupling $f_0$ is larger than the matter couplings in the inflaton sector),
the attractor corresponds to $z^2\approx 6\xi_S s^2 $, which is
the valley along which the squared term proportional to $f_0^2$ in $V_E$, eq.\eq{VE}, nearly vanishes.
Then the potential simplifies to $V_E \simeq (\bp^2/\xi_S s^2)^2 V$, reproducing the
situation considered in~\cite{agravity} and in section~\ref{generic}.  The inflationary predictions are
\beq n_s\approx 1-\frac{2}{N}\stackrel{N\approx 60}{\approx}0.967,\qquad
r\approx \frac{8}{N}\stackrel{N\approx 60}{\approx} 0.13
. \eeq
The scalar amplitude $P_R=M_s^2 N^2/6\pi^2 \bp^2$
is reproduced for  $M_s\approx 1.4 \times 10^{13}\GeV$.

\item {\bf Scalar graviton  inflation}.
In the opposite limit, $M_s\gg M_0$ (obtained when the agravity coupling $f_0$ is smaller than the matter couplings),
the attractor solution  approximately corresponds to a Planckion $s$ frozen at its VEV.
Thereby the Planck constant remains fixed, and the inflaton is $z$, the scalar component of the graviton.
In this limit we obtain Starobinsky inflation~\cite{Starobinsky} that predicts the same $n_s$ as in the previous case
and a smaller value of $r$:

\beq n_s \approx 1 - \frac{2}{N}\stackrel{N\approx 60}{\approx}  0.967,\qquad r \approx \frac{12}{N^2}\stackrel{N\approx 60}{\approx}  0.003.\eeq
The scalar amplitude $P_R = f_0^2 N^2/48\pi^2$
is reproduced for $f_0 \approx 1.8 \times 10^{-5}$.

\item {\bf Higgs inflation}.  We find that, for any value of $M_0/M_s$,
inflation is never dominated by the Higgs, because its quartic  self-coupling $\lambda_H$
(assumed to be positive) is unavoidably larger than the other scalar couplings, taking into account its RG running.
Even assuming that the Higgs has a dominant initial vacuum expectation value~\cite{Bezrukov:2007ep},
in our multifield context inflation starts only after that the field evolution has reached the attractor solution along which $h$ is subdominant,
as exemplified in fig.~\ref{infatt}b.

\end{enumerate}
Notice that in both limits 1.\ and 2.\ the predictions do not depend on $\xi_S$ nor on $\xi_H$.

We next proceed to numerically compute the inflationary predictions corresponding to the intermediate cases by using  the general formulae presented in section \ref{multifield inflationary predictions}.

\begin{figure}[t]
\begin{center}
$$\includegraphics[width=0.45\textwidth]{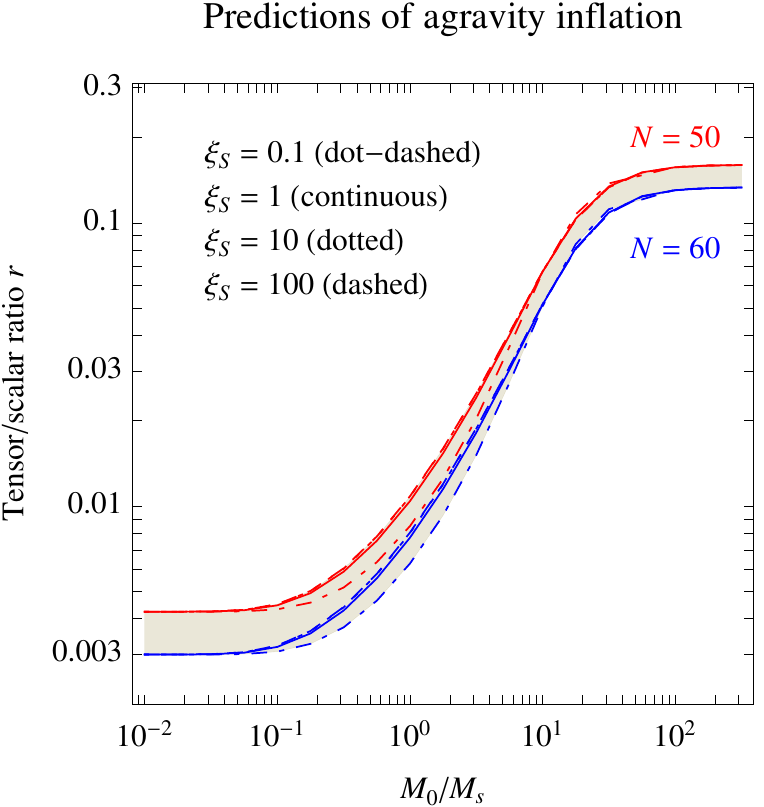}\qquad
\includegraphics[width=0.46\textwidth]{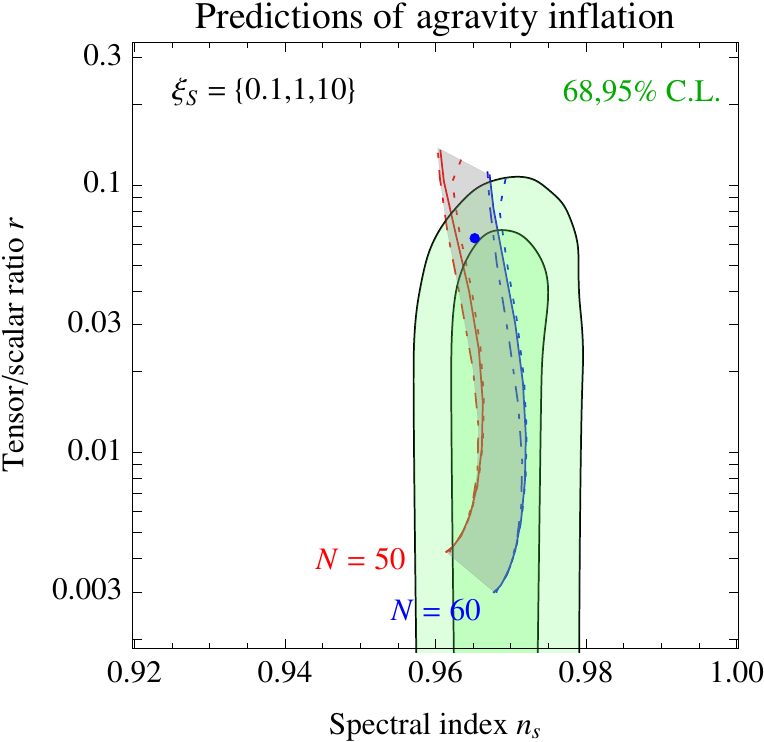}$$
\caption{\em {\bf Left:} predictions for the tensor/scalar ratio $r$ after $N=50$ or $60$
$e$-folds of inflation for various values of $\xi_S$ as function of
$M_0/M_s$.  In the limit where this ratio is large (small), inflation is dominated by the Planckion (the scalar component of the graviton).
{\bf Right:} predictions for the scalar
spectral index $n_s$ and for the tensor/scalar ratio $r$ with the same coding.
The green area is favoured by a global fit of {\sc Planck, BICEP2/Keck}~\cite{BICEP2/Keck:2015tva,Ade:2015fwj,Planck2015}
\label{infpred}}
\end{center}
\end{figure}

Fig.~\ref{infpred} (left) shows the prediction for $r$ at $N =50$ and 60 $e$-folds
while $M_s/M_0$ is varied from small to large values: we find that $r$ smoothly
interpolates between the two limiting cases,
$0.003 < r < 0.13$.
The intermediate region remains negligibly dependent on the value of $\xi_S$.
Furthermore, the value of $n_s-1$ approximately scales as $1/N$ and
remains close to its common value achieved in the two limiting cases.
Fig.~\ref{infpred} (right) shows the prediction in the $(n_s, r)$ plane.\footnote{When both fields are relevant,
our prediction for $(n_s, r)$ lies in the `forbidden region' according to the claim in~\cite{1/N} that assumes single field inflation. 
Unlike in the previous section, all couplings are here small.
Other potentials that lead to similar intermediate values of $r$ are considered in~\cite{attractors}.

}
The prediction is compatible with  the region (in green) preferred by data at 68,  95\% confidence level according to the
latest combination from  {\sc Planck, BICEP2/Keck}~\cite{BICEP2/Keck:2015tva,Ade:2015fwj,Planck2015}.
Next generation experiments could probe $r$ down to $\hbox{few} \times 10^{-3}$.

\section{Cosmology after inflation}\label{reheating}
We here outline the main possibilities for cosmology after inflation in the present context,
and the possible connections with leptogenesis and Dark Matter.
In section~\ref{PRMh} we return to the Higgs mass hierarchy problem.

\subsection{Reheating}
We assumed that the inflaton sector that generates the Planck scale is very weakly coupled to the SM sector,
such that the weak scale is naturally much lighter than the Planck scale.
Because of this, we need to study with special attention how the SM sector can be reheated by the inflaton decays.
The decay of the inflaton $I$ with mass $M_I$ and width $\Gamma_I$ reheats the universe up to a temperature
\begin{equation}
T_{\rm RH}= \left[ \frac{90}{\pi^2 g_*}\,\Gamma^2_I  \bp^2 \right]^{1/4},
\end{equation}
where $g_* \sim 100$ is the number of relativistic degrees of freedom at $T_{\rm RH}$.
We need to compute the total inflaton decay width $\Gamma_I$ and its decay channels,
in order to check if the SM sector is reheated up to a large enough temperature.

\smallskip

Section~\ref{generic} identified the inflaton as the Planckion $s$ and section~\ref{agravity} added
the scalar graviton as a possible extra candidate, finding that the inflaton $I$ is a combination of the two.
We can treat these possibilities jointly given that they have similar couplings, as we now discuss.
The Planckion and the scalar component of the graviton  respectively couple to
\beq \frac{\partial_\mu \mathscr{D}_\mu}{\bp/\sqrt{\xi_S}}\qquad \hbox{and}\qquad \frac{T_{\mu \mu}}{\bp},\eeq
where $T_{\mu\mu}$ is the trace of the energy-momentum tensor and $\partial_\mu \mathscr{D}_\mu$ is the divergence of the dilatation current $\mathscr{D}_\mu = T_{\mu\nu} x_\nu + \mathscr{D}'_\mu$.
According to the Noether procedure, the first term comes from the transformation of the coordinates
$\delta x_\nu\propto x_\nu$; the second term comes from the variations of the fields and
produces a difference between $\partial_\mu \mathscr{D}_\mu$ and $T_{\mu\mu}$.
The theory is explicitly dimensionless in the Jordan frame such that a non-zero divergence arises only at loop level, and is given by
\beq \partial_\mu \mathscr{D}_\mu =
 \frac{\beta_{g_1}}{2g_1 }  Y_{\mu\nu}^{2}+
 \frac{\beta_{g_2}}{2g_2}  W_{\mu\nu}^{2}+ \frac{\beta_{g_3}}{2g_3 }  G_{\mu\nu}^{2}
+\beta_{y_t} H Q_3 U_3+ \beta_{\lambda_H} |H|^4 +\cdots,\label{eq:DD}
\eeq
where $\cdots$ denotes terms beyond the SM and $\beta_g = dg/d\ln\bar\mu$ is the $\beta$ function of the coupling $g$.
The dominant gluon term gives
\beq \Gamma(I\to gg) \approx \frac{|\xi_S| g_3^4 M_s^3}{(4\pi)^5\bp^2}.\eeq
This unavoidable decay channel alone is able of reheating the universe up to
\beq \label{eq:TRH}
T_{\rm RH} \approx 10^7 \GeV \bigg(\frac{M_s}{10^{13}\GeV}\bigg)^{3/2}\eeq
for $\xi_S\sim 1$.

\smallskip

The trace of the energy-momentum tensor
$T_{\mu\mu}$ receives the same loop contribution.
However, it also receives a new, possibly dominant, tree-level contribution.
Indeed, $T_{\mu\mu}$ would vanish in a conformal theory;
we are instead considering a non-conformal theory where the $\xi$ couplings of scalars are generically different from the conformal value $\xi=-1/6$,
as already discussed around eq.~(\ref{eq:Einstein}).
Focusing on the Higgs boson $h$,
the resulting tree-level decay is best computed by transforming the $[(\partial_\mu h)^2 - \xi_H h^2 R]/2$ part of the Lagrangian
 to the Einstein frame
and to canonically normalised fields $h_E  = h \times \bp/ \sqrt{f}$ and $s_E$.
We find the effective operator\footnote{A contribution to it was computed in~\cite{Csaki}.}
\beq -(1+6 \xi_H) \sqrt{\frac{\xi_S}{1+6\xi_S}}  \frac{h^2_E}{2}  \frac{\partial^2 s_E}{\bp}\eeq
that produces a tree-level contribution to the decay width
\beq \Gamma(I \to h_E h_E) =\Gamma(I  \to ZZ) = \frac12 \Gamma(I  \to WW) \approx \frac{(1+6\xi_H)^2 |\xi_S|}{|1+6\xi_S|}
\frac{M_I^3}{64\pi \bp^2}. \eeq
The decays to electroweak vectors arise because their longitudinal components are the Goldstone
components of the Higgs doublet $H$.
For $\xi_{S,H}\sim1$ this channel gives a $T_{\rm RH}\approx 10^9\GeV$, two orders of magnitude larger than in eq.\eq{TRH}.
However, in section~\ref{PRMh} we will find that naturalness of the Higgs mass favours a $\xi_H$ so close to $-1/6$
that $\Gamma(I\to h_Eh_E)$ becomes subdominant with respect to $\Gamma(I\to gg)$.

\bigskip

\subsection{Dark Matter}

So far we neglected possible inflaton decays into the inflaton sector.
For example the minimal model introduced in section~\ref{minimalmodel} contains an extra scalar and an extra fermion.
Such decays are not kinematically allowed if the inflaton is the lightest component of its sector.
This is the case in the minimal model, and likely holds more in general, given that
the Planckion is the light pseudo-Goldstone boson of scale invariance.

However, the inflaton sector must contain fermions in order to provide a negative Yukawa contribution
to the $\beta$ function of the Planckion quartic.
Fermions $\psi$ have an associated $\psi\to - \psi$ symmetry that keeps the lightest fermion stable.

If the lightest fermion within the inflaton sector has no gauge interactions, then it can also couple to the SM sector
behaving as a right handed neutrino $N$.
It can generate the observed light neutrino masses via $N LH$ Yukawa couplings~\cite{seesaw}
(in such a case $M_N\circa{<}10^7\GeV$ in order to avoid an unnaturally large quantum correction to the Higgs mass~\cite{Vissani,FN})
and it can provide baryogenesis via leptogenesis~\cite{leptog}.

\smallskip

The lightest fermion in the inflaton sector is instead a stable Dark Matter candidate
if it cannot couple to the SM sector (for example because it has gauge interactions under the inflaton sector).
Assuming that it is light enough to be produced by inflaton decays,  it inherits
the observed primordial adiabatic density perturbations.\footnote{An alternative production mechanism
can operate even if the fermions are so heavy that decays into them are not kinematically allowed;
however, this alternative  would lead to Dark Matter with isocurvature perturbations~\cite{Tackev}, which are no longer compatible with observations.}

The decay width of the inflaton into Dark Matter  could be computed along the lines of eq.\eq{DD}.
A less model-dependent possibility is that such DM fermions get their mass from another source, independent from the Planckion,
that also breaks scale invariance but at a lower energy scale.
Then such fermion masses would contribute to $\partial_\mu \mathscr{D}_\mu $ and to $T_{\mu\mu}$
as $M \bar\Psi\Psi$
(we are considering, for example, a Dirac mass term)
giving the contribution\footnote{The analogous contribution from SM fermion masses is negligible,
and the Higgs mass terms $M_h$ would give a contribution suppressed by two extra powers
of $M_h/M_I$.}
\beq \label{eq:Gammaf}\Gamma(I \to \bar \Psi \Psi) \approx \frac{ |\xi_S|M^2 M_I}{8\pi(1+6\xi_S) \bp^2}.\eeq
By identifying  the fermion $\Psi$ with Dark Matter, its
abundance is
\beq \Omega_{\rm DM} \equiv \frac{\rho_{\rm DM}}{\rho_{\rm cr}} =\frac{s_0  M}{3H_0^2/8\pi G_N}
 \frac{\Gamma(I \to {\rm DM})}{\Gamma(I \to {\rm SM})}\approx
\frac{0.110}{h^2} \times  \frac{M}{0.40\eV} \frac{\Gamma(I \to {\rm DM})}{\Gamma(I \to {\rm SM})},
\eeq
having inserted the present entropy density ($s_0 = g_{s0} T_0^32\pi^2/45$ with $g_{s0} = {43}/{11}$),
the present Hubble constant $H_0 = h\times 100\,{\rm km}/{\rm sec~Mpc}$,
and the present temperature $T_0 = 2.725\,{\rm K}$.
The observed DM abundance is reproduced for
\beq M\approx (10-200)\TeV \left( \frac{M_I}{10^{13}\GeV}\right)^{2/3}.\eeq
where the lower (higher) estimate applies if $\Gamma(I\to gg)$ ($\Gamma(I\to h_Eh_E)$) dominates.
The proximity of this mass with the range favoured by the hypothesis that DM is a thermal relic is an accident:
this DM candidate leads to negligible non-gravitational signals in agreement with present observations.

\medskip

Finally, we recall that the Higgs potential might be unstable at  energies above $10^{10}\GeV$~\cite{HiggsI}.
The compatibility of this possible instability with cosmology is discussed in~\cite{RUMA}.

\subsection{Inflation and the weak scale}\label{PRMh}
In models that allow for super-Planckian field variations,
inflation becomes a generic natural phenomenon~\cite{Guth:1980zm};
however, small parameters are needed to get the observed amplitude of scalar perturbation
$P_R\approx 10^{-9}$ rather than $P_R \sim 1$ from chaotic inflation.
We obtain a naturally small $P_R$ because we are considering a
dimensionless theory with small couplings.
Small couplings are also needed in order to keep the Higgs mass $M_h$ naturally much smaller than $\bp$.
These models predict a non-trivial connection of the form
\beq  M_h \sim P_R\bp\label{eq:rel}\eeq
between the weak scale, the Planck scale and the amplitude of inflationary perturbations.
Eq.\eq{rel} ignores loop factors $(4\pi)^2$, $e$-fold factors $N\approx 60$ and the possibility that different couplings have different size.
In the rest of the section we add such factors showing that the observed $M_h/\bp$ and $P_R$ are  compatible.

\medskip

As discussed above,  inflation wants $f_0 \circa{>}10^{-5}$.
Thereby, we need to compute the maximal  value of $f_0$ naturally allowed by the Higgs mass.

Whatever sector dynamically breaks scale invariance, it must provide an effective Einstein term $-\bar M_{\rm Pl}^2 R/2$.
Then, inserting such a term in gravitational loop corrections to the Higgs propagator, one estimates a correction to the Higgs mass of order
\beq \label{eq:stima}
\delta M_h^2 \sim \xi_H \frac{g^2(M_g)
  M_{g}^2}{(4\pi)^2} ,\eeq
where, as anticipated in section~\ref{generic},
we assume a soft-gravity scenario where new physics at $M_g$ stops
the growth of the gravitational coupling $g(E)  = E/\bp$. Here we are neglecting the contributions to $\delta M_h^2$ proportional to powers of $M_h^2$ because they are not dangerous from the point of view of naturalness.
The one-loop effect in (\ref{eq:stima}) is proportional to $\xi_H$,  the $\xi$-coupling that provides  interactions between the Higgs and gravity, which do not involve the Higgs mass or derivatives on the external Higgs field.
If $\xi_H$ vanishes, the correction to $M_h^2$ arises at two loops.

\medskip

Within agravity $M_g\sim M_0,M_2$ and the estimate in eq.\eq{stima}  gets replaced by a precise  result~\cite{agravity}:
the  log-enhanced correction to the Higgs mass is described by the following RG equation, valid
at energies between $\bar M_{\rm Pl}$ and $M_g$:
\beq
(4\pi)^2\frac{d  }{d\ln\mub} \frac{M_h^2}{\bar M_{\rm Pl}^2}= -\xi_H [5f_2^4+f_0^4(1+6\xi_H)]+\cdots,\label{eq:RGEm}
\eeq
where $\mub$ is the $\overline{\rm MS}$ renormalization scale and  $\cdots$ denotes negligible terms.
If $\xi_{S,H}\sim 1$, naturalness implies $f_0 \circa{<} 10^{-8}$.
This bound can be relaxed by performing a more complete discussion at the light of the RGE for $\xi_H$~\cite{agravity}:
\beq
(4\pi)^2 \frac{d\xi_H}{d\ln\mub} = (1+6\xi_H)\left(y_t^2-\frac34 g_2^2 - \frac{3}{20} g_1^2+2\lambda_H\right)+\frac{f_0^2}{3}\xi_H(1+6\xi_H)(2+3\xi_H) - \frac53 \frac{f_2^4}{f_0^2}\xi_H
+\cdots, \nonumber
\eeq
where $\cdots$ denote beyond-the-SM terms.
We see that the $\xi$ couplings can naturally acquire two special values:
\begin{itemize}
\item close to zero, $\xi\sim y^2/(4\pi)^2$, where $y$ is a generic coupling of the theory, e.g.\ $y\sim y_t$ in the SM;
\item close to the conformal value,  $\xi + 1/6\sim f_2^4/(4\pi f_0)^2$.
\end{itemize}
In the latter case, the larger $ f_0\sim 10^{-5}$ called by inflation
becomes naturally allowed.
We thereby conclude that the smallness of $M_h/\bp$ and the smallness of $P_R$ are mutually compatible, although
the $\xi$ terms need to lie in special natural ranges.

\section{Conclusions}\label{conclusions}
We  computed the inflationary predictions of models where
mass scales (in particular the Planck scale)
are absent at tree-level and generated only by quantum corrections from dimensionless dynamics.
The same sector that generates the Planck mass does also provide
cosmological  inflation with super-Planckian field variations: the slow-roll parameters are related to the $\beta$-functions of the theory.
We consider two scenarios: single field inflation in an effective field theory approach,  and
 inflation in agravity, which is a UV completion of general relativity coupled to the SM fields.

In the first case, in the limit where couplings are small enough that
gravity effects can be ignored in the Einstein frame,
we showed  the consistency between computations in the Jordan and in the Einstein frames,
obtaining an  inflaton potential which is approximatively quadratic with a cubic correction, leading to
\beq n_s \approx 0.967, \qquad r \approx 0.13.\eeq
If, instead, $\xi_{S}$ and other couplings are large
one can obtain smaller values of $r$, down to about 0.04,
as shown in fig.~\ref{fig:r:vs:ns:ind:Planck:mass}.

\medskip

In the second case, we considered agravity,  the dimensionless extension of Einstein gravity that allows,
for small enough couplings, the natural co-existence of the small weak scale with the much larger Planck scale $\bp$
even in absence of new physics at the weak scale.
Agravity implies an extra scalar component of the graviton
and an extra spin-2 ghost-like graviton that make
quantum gravity renormalizable and controlled by two dimensionless couplings
$f_0, f_2$.
A small amplitude of scalar perturbations is also predicted because this sector
must contain small couplings  such that the weak scale $M_h/\bp$ is naturally small,
resulting in a relation that scales as $P_R \sim M_h/\bp$, more precisely discussed in section~\ref{PRMh}.
We computed the inflationary predictions finding the result in fig.~\ref{infpred} (right):
\beq n_s \approx 0.967, \qquad 0.003 < r < 0.13,\eeq
which agrees with present observations.
Fig.~\ref{infpred} (left) shows that the upper range of $r$ is realised if the `Higgs of gravity' is the lighter scalar that dominates inflation
(this limit gives the effective field theory scenario discussed above);
the lower range is realised if instead the lighter inflationary field is the scalar component of the graviton.

\smallskip

Both in the effective field theory and in agravity, the inflaton decays via Planck-suppressed interactions
(it couples to a combination of the trace of the energy momentum tensor and of the divergence of the dilatation current)
producing a reheating temperature $T_{\rm RH}\sim 10^{7-9}\GeV$.

Furthermore the inflation sector must contain fermions that either behave as right-handed neutrinos (if they have no gauge interactions) or are stable.
In the latter case, they might be light enough that the inflaton can decay in them, providing the observed Dark Matter abundance with adiabatic primordial inhomogeneities
if their mass is around $10-200\TeV$.

\medskip

A  determination of the tensor-to-scalar ratio $r$ in the future can help us to discriminate between different
paradigms behind  inflation and to test quantum theories of gravity such as agravity.

\subsubsection*{Acknowledgments}
We thank   Juan Garc\'ia-Bellido,  Mario Herrero-Valea and Hardi Veerm\"ae for useful discussions.
This work was supported by grants PUT799, MTT8, MTT60, IUT23-6, CERN+, and by EU through the ERDF CoE program. The work of Alberto Salvio has been also supported by the Spanish Ministry of Economy and Competitiveness under grant FPA2012-32828, Consolider-CPAN (CSD2007-00042), the grant  SEV-2012-0249 of the ``Centro de Excelencia Severo Ochoa'' Programme and the grant  HEPHACOS-S2009/ESP1473 from the C.A. de Madrid.

\appendix

\small

\section{RGE and the compatibility of  frames}\label{app}
In this Appendix we give more details about the compatibility between computations performed in the Jordan and Einstein frame in the effective field theory approach described in Section \ref{generic}.
This is particularly relevant since we have to make sure that the RG running in the Jordan frame and in the Einstein frame is consistent.

It has been shown that the results computed in the Jordan and Einstein frames are compatible with each other \cite{George:2013iia}. However, according to the problem at hand, computations can be easier in one frame or another. Usually it is easier to compute the RGEs in the Jordan frame, while inflationary calculations are more simply performed using standard formul\ae{} valid in the Einstein frame. The RGEs of the Jordan frame can be applied to Einstein frame once we keep in mind that also the renormalisation scale has to rescale under a Weyl transformation \cite{George:2013iia}.  In order to fully recover the equivalence one would need to include gravitational loops, that we neglect. 

Let us consider the minimal model described by the Lagrangian (\ref{eq:Jordan:Lagrangian}) in the Jordan frame. Assuming the possibility to neglect quantum gravity corrections eq. \eqref{eq:Ebound}, the one-loop level $\beta$-functions for the couplings in the Jordan frame are given by
\begin{align}
  16 \pi^{2} \beta_{\lambda_{S}} &= 18 \lambda_{S}^{2} + \frac{1}{2} \lambda_{S\sigma}^{2}
  + 4 \lambda_{S} y_{S}^{2} - 4 y_{S}^{4},
  \label{eq:blambda}
  \\
  16 \pi^{2} \beta_{\lambda_{\sigma}} &= 18 \lambda_{\sigma}^{2} + \frac{1}{2} \lambda_{S\sigma}^{2}
  + 4 \lambda_{\sigma} y_{\sigma}^{2} - 4 y_{\sigma}^{4},
  \\
  16 \pi^{2} \beta_{\lambda_{S\sigma}} &= 4 \lambda_{S\sigma}^{2}
  + 6 \lambda_{S\sigma} (\lambda_{S} + \lambda_{\sigma})
  + 4 \lambda_{S\sigma} ( y_{S}^{2} + y_{\sigma}^{2}) -24 y_{S}^{2} y_{\sigma}^{2},
  \\
  16 \pi^{2} \beta_{y_{S}} &= 4 y_{S} (y_{S}^{2} + y_{\sigma}^{2}),
  \\
  16 \pi^{2} \beta_{y_{\sigma}} &= 4 y_{\sigma} (y_{S}^{2} + y_{\sigma}^{2}),
  \\
  16 \pi^{2} \beta_{\xi_{S}} &= \left( \xi_{S} + \frac{1}{6} \right) \left( 6 \lambda_{S} + \frac{1}{6} y_{S}^{2} \right). \label{eq:beta:xiS}
\end{align}
We used the PyR@TE package \cite{Lyonnet:2013dna} to derive the RGEs for the scalar and Yukawa couplings and Ref. \cite{Herranen:2014cua} for the non-minimal coupling $\xi_{S}$.
On the other hand, in the approximation of weakly coupled gravity, the RGEs for the parameters of the Einstein frame Lagrangian \eqref{eq:Einstein:Lagrangian} are
\begin{align}
  16 \pi^{2} \beta_{\Lambda} &= \frac{1}{2} m_{\sigma}^{4}- m_{\psi}^{4},
  \\
  16 \pi^{2} \beta_{\lambda_{\sigma}} &= 18 \lambda_{\sigma}^{2}
  + 4 \lambda_{\sigma} y_{\sigma}^{2} - 4 y_{\sigma}^{4},
  \\
  16 \pi^{2} \beta_{m_{\sigma}^{2}} & = (6 \lambda_{\sigma}  + 4 y_{\sigma}^{2}) \, m_{\sigma}^{2} - 24 y_{\sigma}^{2} m_{\psi}^{2},
  \\
  16 \pi^{2} \beta_{m_{\psi}} &= 4 y_{\sigma}^{2} m_{\psi},
  \\
  16 \pi^{2} \beta_{y_{\sigma}} &= 4 y_{\sigma}^{3},
\end{align}
where the connection between the Jordan and Einstein frame parameters is given by
\begin{eqnarray}
\Lambda &=&  \frac{1}{4} \frac{\lambda_S}{\xi_S^2} \bp^4, \\
  m_{\sigma}^2  &=& \frac{1}{2}  \frac{\lambda_{S\sigma}}{\xi_{S}} \bp^{2}, \\
  m_{\psi} & = &  \frac{y_{S}}{\sqrt{\xi_{S}}} \bp.
\end{eqnarray}
 We see that the RGEs are not perfectly matching with each other. This is because we neglect gravitational corrections 
due to the mixing of gravitational and scalar degrees of freedom in the transformation between the frames.
To match the RGEs of the Jordan frame and the RGEs of Einstein frame in the weak gravity limit, we need to impose 
\begin{equation}
\lambda_{S} \ll \lambda_{S \sigma} \ll \lambda_{\sigma}, \quad \beta_{y_{S}} \ll y_{S}, \quad \beta_{\xi_{S}} \ll \xi_{S},
\label{eq:good:limit}
\end{equation}
on the Jordan frame parameters.
However, since physical observables are frame independent \cite{George:2013iia}, the quantum gravity corrections in the Einstein frame must be reproduced by scalar loops in the Jordan frame.

\footnotesize


\end{document}